\documentclass[utf8]{article-preprint}
\usepackage{url,hyperref,microtype,subcaption}
\usepackage[onehalfspacing]{setspace}
\usepackage{booktabs}

\usepackage{tikz}
\usetikzlibrary{shapes.geometric, arrows.meta, positioning}
\usepackage{xcolor}

\def\keyFont{\fontsize{8}{11}\helveticabold }
\def\firstAuthorLast{Noventa {et~al.}} 
\def\Authors{Nicola Noventa\,$^{1,2}$, Giorgio Daneri\,$^{1}$,  Zhixin Lu\,$^{1,*}$, Matthias Hoelzl\,$^{1}$}
\def\Address{$^{1}$Max Planck Institute for Plasma Physics, Boltzmannstr. 2,  Garching, 85748, Germany \\
$^{2}$Politecnico di Milano, Milan, Piazza Leonardo da Vinci, 32, 20133 Milano MI, Italy }
\def\corrAuthor{Zhixin Lu}

\def\corrEmail{zhixin.lu@ipp.mpg.de}

\begin{document}
\onecolumn
\firstpage{1}

\title {Heterogeneous Acceleration of TRIMEG++ code using OpenMP for Gyrokinetic Simulations}

\author[\firstAuthorLast ]{\Authors} 
\address{} 
\correspondance{} 

\extraAuth{}

\maketitle

\begin{abstract}
Turbulent transport largely determines the performance of magnetic-confinement fusion devices such as tokamaks, and gyrokinetic particle-in-cell (PIC) simulations are a principal tool to predict it. At every time step, tens to hundreds of millions of computational particles (markers) are advanced in the electromagnetic fields, their charge and current are deposited on a mesh, and large sparse linear systems are solved for the new fields, which makes whole-device simulations among the most expensive in plasma physics. Graphics processing units (GPUs) can accelerate this workload substantially, but porting a complex scientific code to them requires restructuring its parallelism and memory traffic, ideally without maintaining a separate version for each GPU vendor. In this work, the gyrokinetic code TRIMEG-C1, whose unstructured triangular mesh and high-order finite elements cover both the confined core and the open-field-line edge of the plasma, is rewritten in C++ as TRIMEG++ and fully offloaded to GPUs with OpenMP target directives, so that one portable code base runs on both AMD and NVIDIA hardware. The particle-to-grid projection is offloaded in both its standard and field-aligned formulations, the PETSc-based field solver runs on the GPU through the library's backends, and, together with the previously offloaded particle pusher, every stage whose cost grows with the number of markers runs on the GPU. All speedups are reported as node-to-node comparisons, i.e.\ , one GPU node against one full powerful CPU node with 256 cores  (2x AMD EPYC 9754). On the TCV-X21 tokamak reference case, the offloaded kernels reproduce the CPU results to a relative accuracy of $10^{-11}$, and the full simulation runs $4.1$--$6.1\times$ faster on a node with two AMD MI300A and $2.1$--$3.0\times$ faster on a node with four NVIDIA A100, with the speedup growing with problem size. The MI300A, whose CPU and GPU share the same memory, needs no explicit data transfers, whereas the A100 requires them to be managed by hand to perform well. These results show that directive-based offloading can accelerate a complex gyrokinetic code across GPU vendors with a single code base, making whole-volume tokamak simulations with hundreds of millions of markers practical on a single GPU node.
\tiny
 \keyFont{ \section{Keywords:} Gyrokinetic, tokamak plasmas, particle simulation, GPU offloading, OpenMP, C++, PETSc} 
\end{abstract}
\section{Introduction}
\label{sec:introduction}
Nuclear fusion offers the prospect of a clean, virtually inexhaustible source of energy, and the central open problem on the path to a fusion reactor is confinement: the performance of a magnetic-confinement device such as a tokamak is largely set by the turbulent transport of heat and particles across the confining field. Because this turbulence cannot be fully diagnosed in experiments, numerical simulation is the principal tool for understanding and predicting it. The reference framework for this task is the gyrokinetic model, in which the fast gyromotion of the charged particles around the magnetic field lines is averaged out analytically, reducing the phase space from six to five dimensions while retaining the drift dynamics and parallel streaming that drive microturbulence \citep{lee1983gyrokinetic,fitzpatrick2014plasma}. Gyrokinetic particle-in-cell (PIC) simulations remain nevertheless extremely demanding: a whole-device run evolves tens to hundreds of millions of marker particles over thousands of time steps, solving large sparse linear systems for the perturbed electromagnetic fields at every step. The majority of existing gyrokinetic codes restrict this calculation to the closed-flux-surface core of the plasma \citep{lin1998turbulent,lanti2020orb5,kleiber2024euterpe}, exploiting a structured mesh and a Fourier decomposition in the poloidal plane; this approach breaks down at the separatrix and cannot be extended across it, yet the scrape-off layer and the divertor govern the exhaust of heat and particles and are essential to predicting plasma-facing component loads in devices such as ITER. Reliable whole-device predictions therefore require a single, self-consistent simulation spanning the confined core and the open-field-line region \citep{hager2022electromagnetic,michels2021genex}, at a computational cost that is among the highest in plasma physics.

TRIMEG-C1 (TRIangular MEsh based Gyrokinetic code) meets this requirement by discretising the poloidal cross-section with an unstructured triangular mesh, which conforms to the separatrix without artificial periodicity and needs no poloidal Fourier filter \citep{lu2019development,lu2024gyrokinetic}. Higher-order $C^1$ finite elements give the accuracy required for electromagnetic simulations across the full plasma volume, and the piecewise field-aligned finite element method (PFAFEM) has been developed on top of this mesh for the efficient simulation of individual toroidal harmonics \citep{Lu2026,lu2025piecewise}. This combination of unstructured geometry and high-order elements makes TRIMEG-C1 more complex, and hence considerably harder to port to a graphics processing unit (GPU), than a structured-mesh core code. 
Over the last decade the growth of supercomputing performance has come almost entirely from massively parallel accelerators rather than from faster CPU cores. The exascale machines that opened the current era, and an increasing share of mid-size academic clusters, deliver most of their throughput and power efficiency through GPUs. For an established scientific code such as TRIMEG-C1, this shift is both an opportunity, a potential order of magnitude in throughput, and a challenge, since exploiting a GPU efficiently requires exposing massive data parallelism and managing the host--device memory traffic explicitly. This must be carried out on hardware from vendors whose native programming models differ substantially, e.g., NVIDIA and AMD. Maintaining one portable code base across this heterogeneous landscape, rather than a separate version per architecture, is one of the central practical problems of current high-performance computing (HPC), and it is the problem this work addresses for TRIMEG-C1.

This paper reports the porting of TRIMEG-C1 to GPUs using OpenMP target directives, a directive-based, vendor-agnostic programming model that keeps a single source code for every platform. The code was first restructured in C++, which both modernises the software and gives OpenMP offloading a uniform compilation path across compiler toolchains. On this base, the two stages that dominate the cost of a time step are offloaded to the device: the particle-to-grid (P2G) projection, in both its non-field-aligned \citep{lu2024gyrokinetic} and field-aligned \citep{Lu2026} formulations, supported by a GPU-native evaluator of the equilibrium B-splines and a spatial sort of the markers; and the PETSc-based field solver, through the library's Kokkos and CUDA GPU backends, with a rank-independent matrix restart format and a solver sub-communicator that decouples the rank count of the solve from that of the particle kernels. Combined with the particle pusher offloaded previously by Daneri \citep{daneri2026openmp}, these bring the whole time-stepping loop onto the device, leaving the host only the MPI reductions, the solver orchestration, and the diagnostics. Everything is developed and benchmarked on production HPC systems: the Viper-GPU cluster at the Max Planck Computing and Data Facility (MPCDF), built with 2 AMD MI300A accelerated processing units (APUs) with $\sim$122.6 TFLOPS/s of peak double precision (FP64), the Raven cluster at MPCDF with 4 discrete NVIDIA A100 GPUs providing $\sim$ 38.8 TFLOPS/s of FP64, and the TOK CPU cluster at IPP Garching, whose nodes, each with two 128-core AMD EPYC 9754 (Zen 4c) processors, serve as the 256-core baseline,  with an estimated peak FP64 performance of 25.4 TFLOP/s, calculated from 2 CPUs$\times$128 cores$\times$3.1 GHz$\times$32 FLOPs/cycle, assuming two 512-bit FP64 FMA instructions per core per cycle. Running the same kernels on a unified-memory APU and on a conventional discrete GPU gives a direct measure of the architecture requirements and performance returns.

The paper is organised as follows. Section~2 summarises the physics model and the numerical methods of TRIMEG-C1 that the rest of the paper builds on. Section~3 describes the GPU offloading itself: the C++ restructuring, the particle pusher, the two particle-projection kernels, the field solver, and their integration into a single GPU-resident time loop. Section~4 reports the performance obtained for each of these components on the TCV-X21 reference case, together with the correctness checks for each kernel. Section~5 validates the fully offloaded code against the CPU reference on the physical observables of an edge-plasma instability. Section~6 summarises the results and discusses their scope and the natural next steps.

\section{TRIMEG-C1 code: physics models and numerical methods}
\subsection{The electromagnetic gyrokinetic model using the mixed variable-pullback scheme}
\label{sec:gk_model}

TRIMEG-C1 models magnetised plasmas in tokamak geometry within the gyrokinetic framework \citep{lee1983gyrokinetic,lee1987gyrokinetic}. The plasma is described by a distribution function $f(z,t)$ defined over the gyrokinetic phase-space coordinates $z=(\mathbf{R},v_\parallel,\mu)$, where $\mathbf{R}$ is the gyrocentre position, $v_\parallel$ the velocity component parallel to the magnetic field, and $\mu=v_\perp^2/(2B)$ the magnetic moment, an adiabatic invariant conserved along particle trajectories in the absence of collisions. In particle-based simulations the $\delta f$ method is used to reduce statistical noise \citep{hatzky2002energy,lu2023full}: the total distribution function is split into a prescribed, time-independent background $f_0$, typically a local Maxwellian, and a time-dependent perturbation $\delta f(z,t)$. The gyrocentre equations of motion decompose correspondingly into an equilibrium part, driven by the unperturbed field, and a perturbed part driven by the fluctuating electromagnetic fields,
\begin{equation}
\dot{\mathbf{R}} = \dot{\mathbf{R}}_0 + \delta\dot{\mathbf{R}}, \qquad
\dot{v}_\parallel = \dot{v}_{\parallel,0} + \delta\dot{v}_\parallel,
\end{equation}
with explicit expressions given by Eqs.~(23)--(26) of \citet{lu2024gyrokinetic}. Since $f_0$ is constant along the unperturbed trajectory, the perturbed marker weight $w_p=\delta f(z_p,t)/g(z_p,t)$, which is the ratio of the perturbed distribution to the marker distribution $g$, evolves according to $\dot w_p = -p_p\,\mathrm{d}[\ln f_0(z_p)]/\mathrm{d}t$, with $p_p=f_0(z_p)/g(z_p)$. \\
Four field equations are solved at every time step in the TRIMEG code \cite{Lu2026}. Only two  are physically independent: the quasi-neutrality equation, whose long-wavelength linearised form reads
\begin{equation}
-\nabla\cdot\Big(\sum_s \frac{q_s n_{0s}}{B\omega_{cs}}\nabla_\perp \delta\Phi\Big) = \sum_s q_s\,\delta n_s ,
\end{equation}
and the parallel component of Amp\`ere's law,
\begin{equation}
-\nabla_\perp^2 \delta A_\parallel = \mu_0\,\delta j_\parallel ,
\end{equation}
for the perturbed potential $\delta\Phi$ and parallel vector potential $\delta A_\parallel$, with $\omega_{cs}=q_sB/m_s$ the cyclotron frequency of species $s$. The perturbed density $\delta n_s$ and parallel current $\delta j_\parallel$ are obtained from $\delta f_s$ by projecting the marker weights onto the finite-element mesh, an operation referred to as particle-to-grid (P2G) throughout this paper and detailed in Section~
\ref{sec:p2g_offload}. The remaining two field equations resolve the cancellation problem \citep{mishchenko2017mitigation,hatzky2019reduction}, a numerical difficulty of electromagnetic gyrokinetic simulations at small electron skin depth. Following the mixed-variable/pullback scheme \citep{mishchenko2014pullback,mishchenko2019pullback,lu2025generalized}, $\delta A_\parallel$ is split into a symplectic and a Hamiltonian part, $\delta A_\parallel = \delta A_\parallel^s + \delta A_\parallel^h$, where $\delta A_\parallel^s$ obeys the ideal Ohm's law $\partial_t \delta A_\parallel^s + \partial_\parallel \delta\Phi = 0$. This decomposition introduces an \textbf{Ohm's law equation} that advances $\delta A_\parallel^s$ in time, while the Amp\`ere equation itself is solved iteratively in the mixed-variable coordinates: a lowest-order solve gives $\delta A_\parallel^h$ by treating the skin-depth term analytically, and an \textbf{iterative Amp\`ere equation} refines it order by order. Together, quasi-neutrality, Amp\`ere, the iterative Amp\`ere correction, and Ohm's law form the complete field solver of TRIMEG-C1; after discretisation each becomes a large sparse linear system, delegated to the PETSc library \citep{petsc-user-ref} and offloaded to the GPU as described in Section~\ref{sec:fieldsolver_offload}.

\subsection{The traditional and the piecewise field-aligned finite element method}
\label{sec:fem_discretisation}

\begin{figure}[htbp]
\begin{center}
  \includegraphics[width=0.55\textwidth]{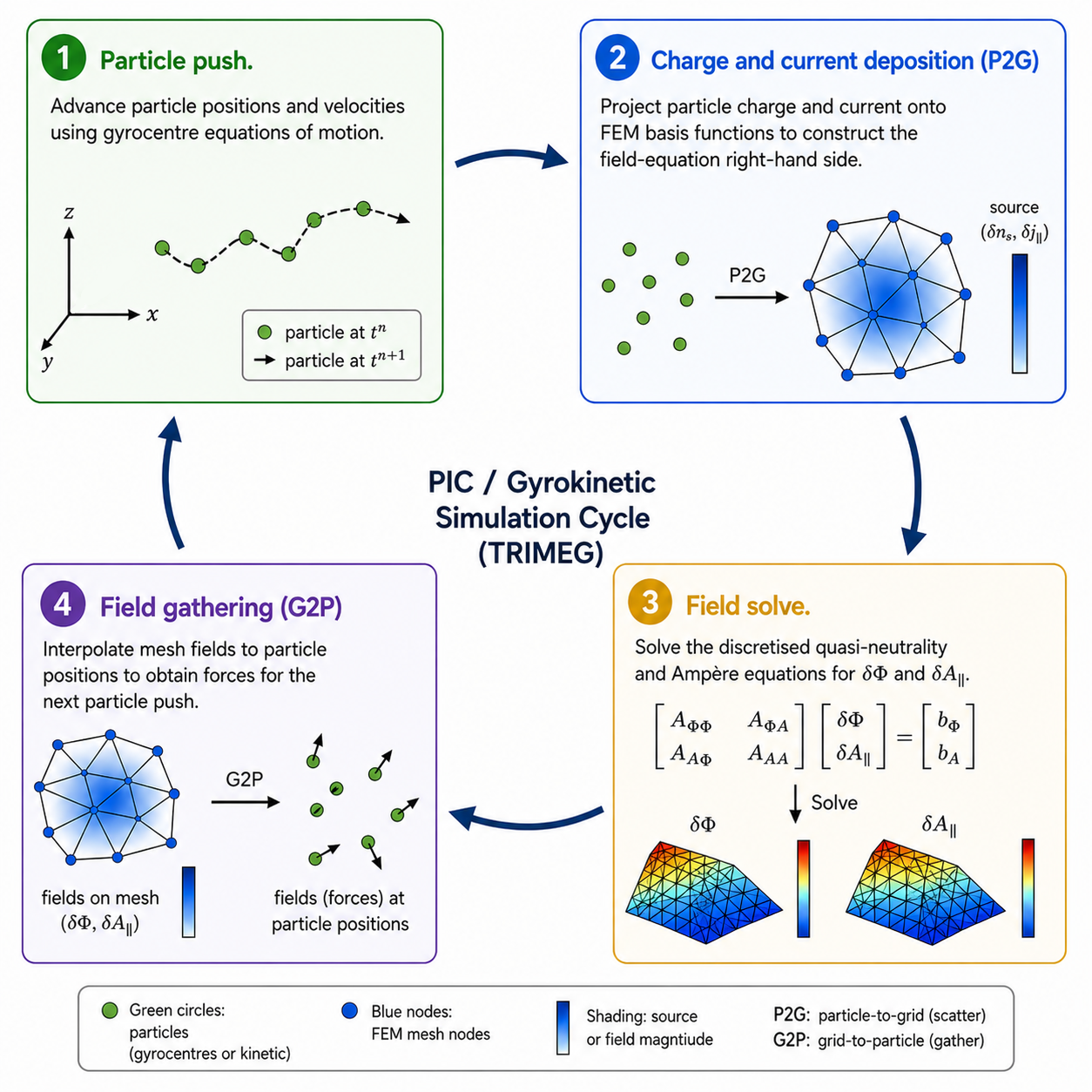}
\end{center}
\caption{The four operations of one PIC step in TRIMEG-C1: particle push, charge/current deposition (P2G), field solve, and field gathering (G2P). This figure was generated with AI for illustration.}
\label{fig:PIC_loop}
\end{figure}

TRIMEG-C1 follows the particle-in-cell (PIC) method \citep{lee1987gyrokinetic,lanti2020orb5}: the markers are tracked in the continuous phase space while the fields are represented on a fixed mesh, and the two descriptions are coupled at every step, as sketched in Figure~\ref{fig:PIC_loop}. The fields on the mesh push the markers by interpolating their values at the marker position in the continuous phase space. Then the markers deposit charge and current onto the mesh (P2G) as the source of the field equations. In configuration space TRIMEG-C1 uses global cylindrical coordinates $(R,\varphi,Z)$ adapted to tokamak geometry; in velocity space it uses $(v_\parallel,\mu)$, the natural coordinates aligned with $\mathbf{B}$ once the gyrophase has been averaged out.

The fields are discretised as a tensor product of a two-dimensional finite-element basis in the poloidal $(R,Z)$ plane, on the unstructured triangular mesh, and a one-dimensional B-spline basis in the periodic toroidal direction. On each triangle, mapped from the physical $(R,Z)$ plane to reference coordinates $(\xi,\eta)$, TRIMEG-C1 uses reduced $C^1$ quintic elements, chosen because the field equations involve second-order derivatives, for which linear ($C^0$) elements are not accurate enough. Each vertex then carries $N_{\mathrm{dof}}=6$ degrees of freedom (the function value and its first and second partial derivatives) for $3\times 6=18$ degrees of freedom per triangle and $6N_{\mathrm{vert}}$ independent poloidal degrees of freedom over the whole mesh, $N_{\mathrm{vert}}$ being the number of vertices \citep{jardin2004triangular}. In the toroidal direction, the domain is periodic and discretised with $N_\varphi$ cubic B-spline basis functions $N_j(\varphi)$, so that a three-dimensional field is represented as
\begin{equation}
Y(R,\varphi,Z) = \sum_{i=1}^{6N_{\mathrm{vert}}}\sum_{j=1}^{N_\varphi} \hat\Lambda_i(\xi,\eta)\,N_j(\varphi)\,\hat Y_{i,j} ,
\label{eq:3d_expansion}
\end{equation}
with $\hat\Lambda_i$ the basis functions in the physical space. Substituting Eq.~\eqref{eq:3d_expansion} into the field equations and applying the Galerkin procedure yields the linear systems whose right-hand side must be reassembled from the current marker distribution at every step.

Equation~\eqref{eq:3d_expansion} can be evaluated in two ways. In the traditional, non-field-aligned scheme, the basis functions are defined directly in the cylindrical coordinates $(R,\varphi,Z)$, and the toroidal sum can be restricted to a single Fourier mode of order $n_c$ when only one toroidal harmonic is of interest \citep{Lu2026}. Tokamak microinstabilities, however, are strongly elongated along the magnetic field ($k_\parallel \ll k_\perp$) \citep{beer1995field}, so resolving them on this grid requires many toroidal degrees of freedom. The piecewise field-aligned finite element method (PFAFEM) \citep{lu2025piecewise,lu2025trimeg,Lu2026} keeps the same poloidal mesh but defines the basis functions in piecewise field-aligned coordinates: each basis function follows the magnetic field line inside a toroidal wedge, so that field-aligned structures are captured with far fewer toroidal planes. The price is that, before the P2G projection or the G2P gather, every marker must be mapped along the field line from $(R,Z,\varphi)$ to the field-aligned position $(R_{\mathrm{FA}},Z_{\mathrm{FA}})$ of the relevant toroidal plane, a mapping evaluated through pre-computed three-dimensional B-splines. Both the non-PFAFEM and the PFAFEM schemes are implemented in TRIMEG-C1, and the GPU offloading of both P2G formulations is presented in Section~\ref{sec:p2g_offload}.
\section{Heterogeneous Acceleration of TRIMEG code}
\label{sec:offloading}
\begin{figure}[htbp]
\centering
\begin{tikzpicture}[
  node distance=0.5cm,
  gpubox/.style={rectangle, rounded corners=4pt, minimum width=7cm,
                 minimum height=1.1cm, align=center,
                 draw=teal!60!black, fill=teal!15, font=\small},
  petscbox/.style={rectangle, rounded corners=4pt, minimum width=7cm,
                   minimum height=0.8cm, align=center,
                   draw=violet!60!black, fill=violet!15, font=\small},
  terminal/.style={rectangle, rounded corners=14pt, minimum width=5cm,
                   minimum height=0.8cm, align=center,
                   draw=gray!60, fill=gray!20, font=\small},
  arr/.style={->, >=stealth, thick, gray!70}
]

\node[terminal] (start) {Start RK4 substep};

\node[gpubox, below=of start] (curr) {
  \textbf{Calculate current}\\
  {\footnotesize P2G: scatter $w_p \cdot v_{\parallel,p}$ onto grid DOFs}
};

\node[petscbox, below=of curr] (ampere) {\textbf{Ampère solver}};

\node[gpubox, below=of ampere] (iteramp) {
  \textbf{Iterative Ampère solver}\\
  {\footnotesize G2P + P2G inside each iteration}
};

\node[gpubox, below=of iteramp] (dens) {
  \textbf{Calculate density}\\
  {\footnotesize P2G: scatter $w_p$ onto grid DOFs}
};

\node[petscbox, below=of dens] (poisson) {\textbf{Quasi-neutrality solver}};

\node[petscbox, below=of poisson] (ohm) {\textbf{Ohm solver}};

\node[gpubox, below=of ohm] (push) {\textbf{Push particles}};

\node[terminal, below=of push] (end) {End RK4 substep};

\draw[arr] (start)   -- (curr);
\draw[arr] (curr)    -- (ampere);
\draw[arr] (ampere)  -- (iteramp);
\draw[arr] (iteramp) -- (dens);
\draw[arr] (dens)    -- (poisson);
\draw[arr] (poisson) -- (ohm);
\draw[arr] (ohm)     -- (push);
\draw[arr] (push)    -- (end);

\node[below=0.6cm of end, xshift=-2cm] (leg1)
  [rectangle, rounded corners=3pt, minimum width=0.4cm,
   minimum height=0.4cm, draw=teal!60!black, fill=teal!15] {};
\node[right=0.15cm of leg1, font=\footnotesize] {GPU kernel (OpenMP offloading)};

\node[below=0.2cm of leg1]  (leg2)
  [rectangle, rounded corners=3pt, minimum width=0.4cm,
   minimum height=0.4cm, draw=violet!60!black, fill=violet!15] {};
\node[right=0.15cm of leg2, font=\footnotesize]
  {GPU via the PETSc backend (Section~\ref{sec:fieldsolver_offload})};

\end{tikzpicture}
\caption{One RK4 substep of the fully GPU-offloaded TRIMEG-C1 time loop.
  Teal boxes are OpenMP \texttt{target} kernels: the particle pusher and the P2G projection
  (Section~\ref{sec:p2g_offload}), which appears three times per
  substep, namely, in the current deposition, inside the iterative
  Amp\`ere solver, and in the charge-density deposition. Violet boxes
  are the field solves, offloaded through the PETSc GPU backend
  (Section~\ref{sec:fieldsolver_offload}). The host retains only the
  MPI reductions of the deposited moments, the solver orchestration
  (Section~\ref{sec:fullsim}), and diagnostics.}
\label{fig:rk4_flowchart}
\end{figure}
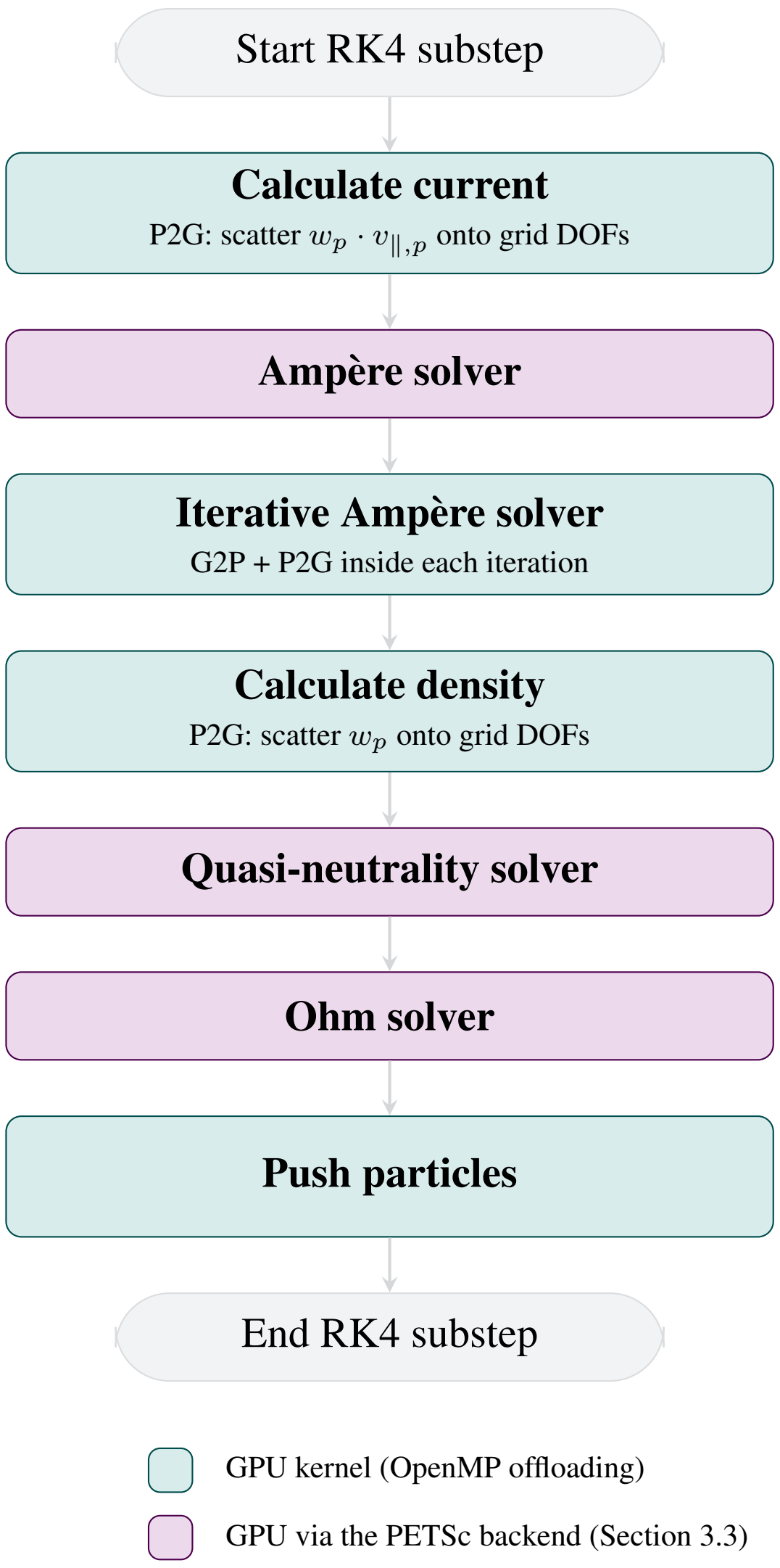

Figure~\ref{fig:rk4_flowchart} shows the sequence of operations performed at
each substep of the fourth-order Runge--Kutta integrator in TRIMEG-C1: the
deposition of the marker moments onto the grid, the solution of the four field
equations of Section~\ref{sec:gk_model}, and the advance of the markers with the fields
gathered at their positions. All of these stages are offloaded to the device
in the present work, leaving the host only the MPI reductions, the solver
orchestration, and the diagnostics. The following sections describe the C++
restructuring on which the port is built
(Section~\ref{sec:cpp_restructuring}), the projection and
field-solver kernels (Sections~\ref{sec:p2g_offload}
and~\ref{sec:fieldsolver_offload}), and their integration into a fully
GPU-resident time loop (Section~\ref{sec:fullsim}). The particle pusher developed by Daneri
will be included in a separate article \citep{daneri2026openmp}. 
Note that since we use a powerful pure CPU computing node for comparison as mentioned in Section~\ref{sec:introduction}, the acceleration ratio of the GPU node looks lower than that in other work where the CPUs were less powerful  \cite{zhang2018heterogeneous,cheng2016gyrokinetic}.

\subsection{Porting TRIMEG from Fortran to C++}
\label{sec:cpp_restructuring}

TRIMEG-C1 was originally written in Fortran, an excellent choice for codes that are dominated by linear algebra operations. It supports multidimensional arrays natively, allowing for a clean and compact syntax. However, all of the above apply to traditional CPU-based systems. Fortran has disadvantages for the TRIMEG project moving towards heterogeneous computing. First and foremost, most GPU programming languages and application programming interfaces (APIs) are native in C++, and a Fortran counterpart does not exist. Among these are HIP, Kokkos, and Alpaka. In contrast, CUDA offers Fortran support with CUDA Fortran, but is not suited to achieve portability across different platforms, e.g., AMD. Portability is a fundamental pillar of the offloading strategy of TRIMEG, since the code needs to be compatible with all available cluster architectures, which include both NVIDIA and AMD GPU hardware. The only possible choices are directive-based APIs that also implement GPU offloading, that is, OpenMP and OpenACC. Support for the latter is less in AMD GPU compilers. Hence, we choose OpenMP as the GPU offloading model.
Nonetheless, OpenMP offloading support in Fortran compilers is still in the development stage, with more support for AMD GPUs/APUs for selected features from the newer standard, and requires extensive debugging to eventually achieve a working solution. This has been thoroughly documented in the previous work about TRIMEG GPU acceleration and portability of \cite{daneri2026openmp}.

Due to several performance and portability constraints encountered, the decision was made to translate the whole codebase to C++. The purpose was to implement a fully GPU-accelerated version of TRIMEG-C1, and has been achieved as described in this paper. The offloading process in C++ has proven to be significantly more streamlined, requiring less debugging and achieving better performance.
The translation was carried out with extensive use of LLMs, even though the process was supervised. No AI agent was used. The structure mimics that of the original Fortran code as closely as possible, both to preserve the model semantics and to decrease the learning curve of Fortran developers. A testing infrastructure has been implemented, consisting of a continuous integration (CI) pipeline with regression and validation tests for each code component. 

The initial design choice was made to keep an external Fortran library implementing B-splines, in order to minimise the discrepancies during the validation steps of the C++ port. This was a temporary measure and results in more complex data management for C++ GPU kernels, which call Fortran subroutines internally. A native C++, GPU-optimised library is implemented to substitute the current one, which has shown to be inefficient on GPU platforms.

\subsection{GPU offloading of the particle projection}
\label{sec:p2g_offload}

The particle-to-grid (P2G) projection of Section~\ref{sec:fem_discretisation}, which deposits the marker moments onto the finite-element basis of Eq.~\eqref{eq:3d_expansion}, is executed for the entire marker population at every substep (Figure~\ref{fig:rk4_flowchart}): once for the current deposition, once inside each iteration of the iterative Amp\`ere solver, and once for the charge-density deposition. Its cost scales linearly with the number of markers and, at the tens to hundreds of millions of markers used in production runs, it is one of the dominant contributors to the time step. It is also well suited to GPU acceleration in principle, since the same sequence of operations is applied independently to every marker, including basis evaluation, coordinate transformation, and scatter. The CPU implementation of P2G already existed in TRIMEG-C1; offloading it to the GPU, in both its non-field-aligned (NFA) and field-aligned (FA) formulations (Section~
\ref{sec:fem_discretisation}), is one of the central contributions of this work, together with the GPU field solver of Section~\ref{sec:fieldsolver_offload} and the integration of Section~\ref{sec:fullsim}.

\subsubsection{Non-field-aligned projection}
\label{sec:nfa_offload}

In the NFA path, the host triangle and local coordinates $(\xi,\eta)$ of every marker are known before the scatter step, so the kernel evaluates the 18 local basis functions at $(\xi,\eta)$, applies the local-to-physical Jacobian transformation, and atomically scatters the weighted contribution $\hat\Lambda_i(\xi_p,\eta_p)\,w_p\,e^{-in_c\varphi_p}$ onto the 18 degrees of freedom of the host triangle. Mesh data, including the sparse Jacobian pattern, the per-triangle Jacobian components, and the triangle--vertex connectivity, is mapped to the device once, after construction. The particle buffers and the complex-valued accumulator, are allocated once and refreshed with \texttt{target update} at every call, avoiding repeated device allocation. The latter is re-implemented as a plain \texttt{double*} array since OpenMP target regions do not support \texttt{std::complex}.

A first port of the CPU loop features a parallel outer loop over the particle markers, while the inner loops over the 18 degrees of freedom (DOFs) are left serial inside each GPU thread; it achieves a speedup of about $3\times$ over the CPU baseline (Section~\ref{sec:perf_nfa}), but profiling with the AMD ROCm profiler on Viper-GPU showed that this version is limited neither by compute throughput nor by memory bandwidth: the three per-thread private arrays needed for the serial inner loops exceed the available vector registers and are spilled to slow scratch memory, limiting the achieved performance to only $23.2$~GFLOP/s, and the hardware occupancy to $28\%$. A second kernel version exposes all three loops over markers, vertexes, and DOFs to the OpenMP runtime with \texttt{collapse(3)}, so that each of the resulting $18\,N_p$ threads computes a single scalar contribution and no per-thread array is needed. This redesign removes the scratch traffic entirely and raises the achieved throughput to $760.6$~GFLOP/s (a $33\times$ improvement) and the occupancy to $50\%$. The gain in kernel time is smaller, a $2.4$--$2.7\times$ reduction, because the two versions do not perform the same amount of work: quantities that the first version evaluated once per marker and reused across all 18 DOFs are now recomputed independently by every thread, so the total number of floating-point operations grows substantially. The redesign thus trades additional, cheap arithmetic for the elimination of expensive scratch-memory accesses, and the net effect brings the speedup over the CPU baseline from about $3\times$ to $7$--$8\times$ (Section~\ref{sec:perf_nfa}).

\subsubsection{Field-aligned projection}
\label{sec:fa_offload}

The field-aligned path is structurally harder: the toroidal B-spline has compact support over four planes; therefore every marker contributes to four toroidal planes, and for each of them it must first be mapped along the magnetic field line to a different reference-plane position $(R_{\mathrm{FA}},Z_{\mathrm{FA}})$ through the equilibrium B-splines \texttt{Rmid3d} and \texttt{Zmid3d}, and then located in the poloidal mesh by a triangle search. The GPU kernel is therefore split into two device kernels that communicate through intermediate buffers: a memory-bound \emph{mapping and search} kernel, launched over the $4N_p$ (marker, toroidal shift) pairs, and an atomic-bound \emph{basis and scatter} kernel, launched with \texttt{collapse(3)} over the additional factor of 18 from the basis functions. Splitting the two stages avoids forcing a single launch configuration and register footprint onto operations with very different performance characteristics.

Two optimisations target the memory-bound mapping kernel specifically. First, a GPU-native C++ evaluator replaces the generic Fortran B-spline routine used elsewhere in TRIMEG-C1. It is specialised to the order-3 splines used for the equilibrium maps, and it expands the de~Boor recursion into closed-form scalar expressions instead of the generic machinery of the Fortran routine, which removes both the per-call work arrays and the register pressure they cause. Here order  is degree plus one, in the convention of the underlying B-spline library, so that these are \emph{quadratic} interpolants and not the \emph{cubic} toroidal basis of Eq.~\eqref{eq:3d_expansion}. An A/B benchmark running the two routines on the same evaluation points, averaged over three passes of $8N_p$ evaluations at $N_p=1$--$16$~million on the MI300A, gives the native evaluator a speedup of $3.9$--$4.6\times$ on the GPU and $8.7$--$12.4\times$ on the CPU, the host figures being single-core; it also accelerates the field-gather step of the particle pusher. Second, an optional GPU counting sort reorders the markers by their host cell in the equilibrium B-spline grid immediately before the mapping kernel, so that neighbouring GPU threads read neighbouring coefficient blocks and follow similar paths through the triangle-search structure. Measured on the MI300A with eight MPI ranks over two GPUs and averaged over nine projection calls, the sort improves the memory-bound mapping kernel by a factor of $2$--$3$ while leaving the atomic-bound scatter kernel essentially unaffected, for a net speedup of the whole projection of $1.5$--$1.6\times$ at production marker counts. It is kept optional because the reordering itself costs $10$--$13\%$ of the projection time at those counts and $22\%$ at $1$~million markers, where it outweighs the gain and makes the sort a net slowdown ($0.84\times$); the sort therefore only pays off above a few million markers. Together, the native evaluator and the spatial sort accelerate the field-aligned projection by a factor of $7$--$9\times$ compared with an equally optimised CPU baseline at production scale (Section~\ref{sec:perf_fa}).

The correctness of both kernels is verified by comparing the GPU output element by element against the CPU reference on the same input: the projected density and particle weights agree to $10^{-12}$--$10^{-11}$ in relative terms, consistent with the different order of floating-point accumulation on the two architectures rather than with any algorithmic discrepancy (Sections~\ref{sec:perf_nfa} and~\ref{sec:perf_fa}).

\subsection{GPU offloading of the field solvers}
\label{sec:fieldsolver_offload}
\subsubsection{Benchmark of the field solver}
\label{sec:solver_bench}
In TRIMEG, the fields are obtained from the particle moments by solving a
set of linear systems on the grid at every time step (Eqs.~(80)--(83) of
\cite{lu2024gyrokinetic}.
All the solver measurements reported in Section~\ref{sec:solver_results} are made with a stand-alone test rather than with the full code, but the numerical studies are carried out with the necessary components of the field solver in the TRIMEG code. The test reads the
same equilibrium and mesh as a production run and assembles its matrices
through the same general assembly routine used for the production field
equations, but with simplified coefficients: the operator it builds is the
field-aligned three-dimensional Laplacian with a constant coefficient,
without the spatially varying equilibrium coefficients of the
quasi-neutrality and Amp\`ere operators. The system is then
solved repeatedly for a manufactured solution, a smooth bell function
whose analytic coefficients are known, so that every solve can be checked
against the exact answer.

The benchmark is therefore a structural test of the production
solves, not an exact reproduction of them. The operators still share the items that determine the iteration cost: the same $C^1$ basis and field-aligned formulation, the
same mesh and number of unknowns, and the same sparsity pattern, that is, the same work per matrix--vector product, which dominates the runtime. What it does not share are the equilibrium-dependent coefficients,
so the iteration counts measured here need not coincide with those of the
production quasi-neutrality and Amp\`ere solves, and the observations concerning the solver robustness
in Section~\ref{sec:solver_results} should be re-checked on the production
operators before being taken as final. Within this scope, the test gives a
clean benchmark: it isolates the solver from the rest of the time step evolution,
uses the production tolerance ($\mathrm{rtol} = 10^{-5}$), and provides a
correctness check on every configuration that was applied.
\subsubsection{The PETSc Library and Its GPU Backend}
\label{sec:solver_petsc}

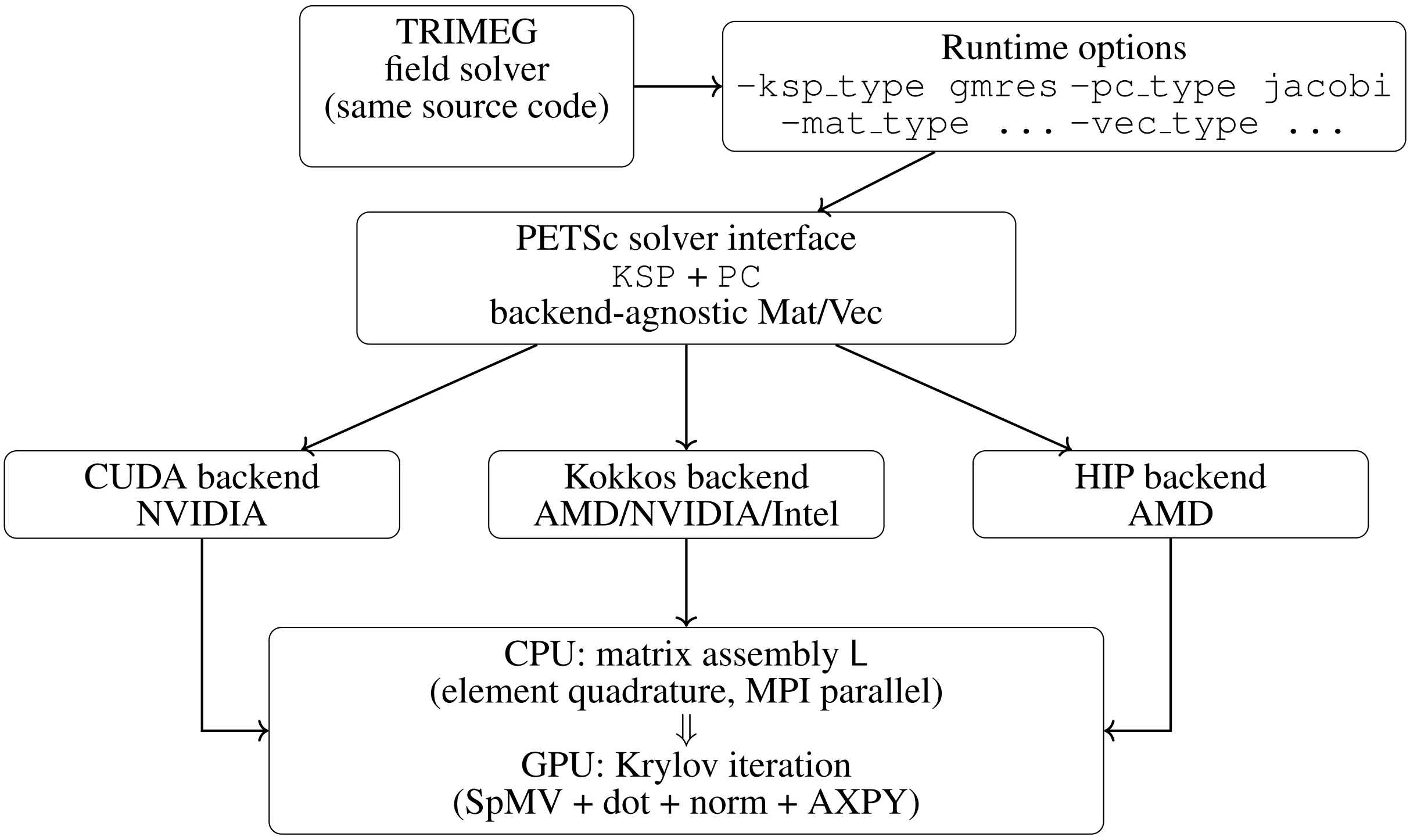
\begin{figure}[t]
\centering
\begin{tikzpicture}[
    node distance=5mm and 10mm,
    box/.style={draw, rounded corners, align=center, minimum height=8mm},
    arrow/.style={->, thick}
]

\node[box, minimum width=38mm] (trimeg) 
{TRIMEG\\ field solver \\(same source code) \\};

\node[box, right=of trimeg, minimum width=55mm] (options)
{Runtime options\\
\texttt{-ksp\_type gmres} 
\texttt{-pc\_type jacobi}\\
\texttt{-mat\_type ...}
\texttt{-vec\_type ...}};

\draw[arrow] (trimeg) -- (options);

\node[box, below=of trimeg, xshift=25mm, minimum width=75mm] (petsc)
{PETSc solver interface\\
\texttt{KSP} + \texttt{PC}\\
backend-agnostic Mat/Vec};

\draw[arrow] (options) -- (petsc);

\node[box, below left=12mm and -5mm of petsc, minimum width=45mm] (cuda)
{CUDA backend\\
NVIDIA};

\node[box, below=12mm of petsc, minimum width=45mm] (kokkos)
{Kokkos backend\\
AMD/NVIDIA/Intel};

\node[box, below right=12mm and -5mm of petsc, minimum width=45mm] (hip)
{HIP backend\\
AMD};

\draw[arrow] (petsc) -- (cuda);
\draw[arrow] (petsc) -- (kokkos);
\draw[arrow] (petsc) -- (hip);

\node[box, below=10mm of kokkos, minimum width=95mm] (execution)
{
CPU: matrix assembly $\mathsf{L}$\\
(element quadrature, MPI parallel)\\
$\Downarrow$\\
GPU: Krylov iteration\\
(SpMV + dot + norm + AXPY)
};

\draw[arrow] (cuda.south) |- (execution);
\draw[arrow] (kokkos.south) -- (execution.north);
\draw[arrow] (hip.south) |- (execution);

\end{tikzpicture}

\caption{Runtime-selectable PETSc solver framework and hybrid
CPU--GPU execution strategy in the TRIMEG field solver. The same executable supports
different solver algorithms and device backends through runtime options.}
\label{fig:petsc_gpu_strategy}
\end{figure}

PETSc (Portable, Extensible Toolkit for Scientific Computation) is the
de-facto standard library for large sparse linear algebra on distributed
machines, and TRIMEG uses it for its field solvers.
Figure~\ref{fig:petsc_gpu_strategy} summarises the PETSc solver framework and the hybrid
CPU--GPU execution strategy adopted for the TRIMEG field solver.

Three of PETSc's design choices shape the offloading strategy. First, the solver algorithm is entirely a runtime choice: TRIMEG calls \texttt{KSPSetFromOptions}, so the Krylov method, preconditioner, tolerances and monitoring are all set from the command line (e.g.\ \texttt{-ksp\_type gmres -pc\_type jacobi}) without the need of recompiling the code. Second, matrices and vectors are backend-agnostic: the standard distributed matrix becomes a GPU matrix by selecting a device type at runtime, either a vendor-native backend (\texttt{cuda} on NVIDIA, \texttt{hip} on AMD) or the Kokkos backend \citep{kokkos}, which compiles unchanged for AMD, NVIDIA, and Intel devices; this work uses Kokkos on Viper and the native CUDA backend on Raven (Section~\ref{sec:solver_results}). Whichever type is selected, PETSc assembles the matrix on the host once and thereafter executes the entire Krylov iteration on the device, so the only host--device traffic per solve involves the right-hand side and the solution. Third, the GPU-supported components are fewer: of the preconditioners considered here, only Jacobi has a device implementation, while the incomplete factorisations (ILU, ICC) are not supported, including the block-Jacobi/ILU(0) combination that is PETSc's CPU default, which is accounted for in the benchmark protocol of Section~\ref{sec:solver_results}.

The workload distribution matches the structure of the problem: matrix assembly, a loop over elements already parallelised over MPI ranks, stays on the CPU, while the iterative solve is offloaded to harness the GPU's High Bandwidth Memory (HBM) bandwidth (e.g.\ $5.3$\,TB/s peak on the MI300A \citep{amd_mi300a_datasheet}), as it consists of hundreds of memory-bandwidth-bound sparse matrix--vector products per time step.

Finally, the PETSc binary format makes the stored operators independent of
the domain decomposition: \texttt{MatView} writes a single file for the
global matrix and \texttt{MatLoad} redistributes its rows over whatever
communicator it is given, whereas the original TRIMEG-C1 checkpoint wrote
one file per rank and could only be reloaded on the same rank count. This is
what lets the solve run on a different number of ranks from the rest of the
simulation (Sections~\ref{sec:solver_results_ranks} and~\ref{sec:fullsim}),
and it also allocates exactly the stored nonzeros instead of an upper-bound
estimate, with the memory saving quantified in
Section~\ref{sec:solver_results}.

\subsection{Integration into a fully GPU-resident time loop}
\label{sec:fullsim}

Bringing the particle pusher \cite{daneri2026openmp}, the two P2G kernels (Section~\ref{sec:p2g_offload}) and the GPU field solver (Section~\ref{sec:fieldsolver_offload}) together into the single time loop of Figure~\ref{fig:rk4_flowchart}, running for thousands of steps rather than in isolation, involves more than combining the three ports: with the pusher and the coordinate updates also executed on the device, every stage whose cost scales with the marker count runs on the GPU, and the host retains only the MPI reductions of the deposited moments, the orchestration of the solver, and the diagnostics.

Two integration-specific mechanisms were needed to reach this point. First, the field solve and the particle kernels require a different number of MPI ranks: the particle kernels get faster with more ranks, while the small, mesh-sized matrices of Section~\ref{sec:fieldsolver_offload} are solved fastest on very few ranks per GPU, and running the solve on every rank of the node makes it slower rather than faster once several ranks contend for the same GPU hardware resources. A \emph{solver sub-communicator} resolves this conflict: a runtime option selects how many ranks take part in the PETSc solve, spread evenly across the GPUs of the node, while the remaining ranks skip the linear algebra entirely; the rank-independent binary restart format of Section~\ref{sec:solver_petsc} makes this possible, since \texttt{MatLoad} works on a communicator of any size. Second, on the discrete A100 GPUs of Raven the code uses explicit device mapping rather than managed memory throughout the time loop, since it proved both more controllable and measurably faster; this requires the mesh and spline data to be mapped once at start-up while the fields and particle arrays are re-mapped at every kernel launch, following the same present-table and raw-pointer rationale already used for the projection kernels of Section~\ref{sec:nfa_offload}. Both mechanisms, and the performance they enable, are quantified in Section~\ref{sec:fullsim_performance}.

\section{Performance analyses of GPU offloading}
\label{sec:performance}
\subsection{The test case and HPCs}
\label{sec:testcase_hpc}
\subsubsection{The TCV-X21 test case and simulation parameters}
\label{sec:tcv_case}
The benchmark case used across all runs is the TCV-X21 configuration ~\citep{oliveira2022tcvx21}, a
well-established reference case for edge-plasma turbulence codes based on
experimental data from the TCV tokamak. This case is based on a benchmark scenario that has been used in various codes such as GENE-X, GRILLIX and TRIMEG-C1 \cite{Lu2026,body2022development,ulbl2023influence}. The simulation
is initialised from an EQDSK equilibrium file providing the poloidal flux function
$\psi_p$ and related profiles such as the temperature and density. The
computational domain covers both closed and open field-line regions, with
the plasma separatrix explicitly included. The unstructured triangular mesh
is aligned with the flux surfaces and features additional refinement near
the X-point; the production mesh used in the present benchmarks consists
of 5037 vertices.

\subsubsection{Hardware and build configuration}
\label{sec:bench_tok}

Three systems are used. TOK, a CPU cluster at IPP Garching, provides the
baseline: each node hosts two AMD EPYC 9754 processors (256 cores total)
and is built with GCC~14 in Release mode. Raven and Viper-GPU provide the GPU
hardware: Raven nodes carry four discrete NVIDIA A100 GPUs alongside two
Intel Xeon Platinum 8360Y processors     (36 cores each, 72 in total, 512~GB DDR4), while
Viper-GPU nodes carry two AMD MI300A APUs, each integrating 24 Zen~4 CPU
cores and a CDNA3 GPU die on 128~GB of shared HBM3 memory (48 cores and
256~GB HBM3 per node), eliminating the discrete CPU--GPU memory boundary
present on Raven. Table~\ref{tab:build_config} summarises the compiler and
key offload flag used on each GPU system; the XNACK support enabled by
\texttt{--offload-arch=gfx942:xnack+} on Viper is what allows unified
memory and page-migration between the CPU and GPU address spaces of the
MI300A.

\begin{table}[htbp]
\centering
\caption{Hardware, compiler, and key OpenMP offload flag on each benchmark system.}
\label{tab:build_config}
\begin{tabular}{llll}
\toprule
Machine & Processor & Compiler & Key offload flag \\
\midrule
TOK & (EPYC 9754 CPU, 2/node) & GCC~14 & - \\
Raven & NVIDIA A100 (4/node) & \texttt{nvc++} & \texttt{-mp=gpu -gpu=cc80} \\
Viper-GPU & AMD MI300A (2/node) & \texttt{amdclang++} & \texttt{--offload-arch=gfx942:xnack+} \\
\bottomrule
\end{tabular}
\end{table}

\subsection{Performance of particle projection: non-field-aligned FEM}
\label{sec:perf_nfa}

\begin{figure}[htbp]
  \centering
  \includegraphics[width=\textwidth]{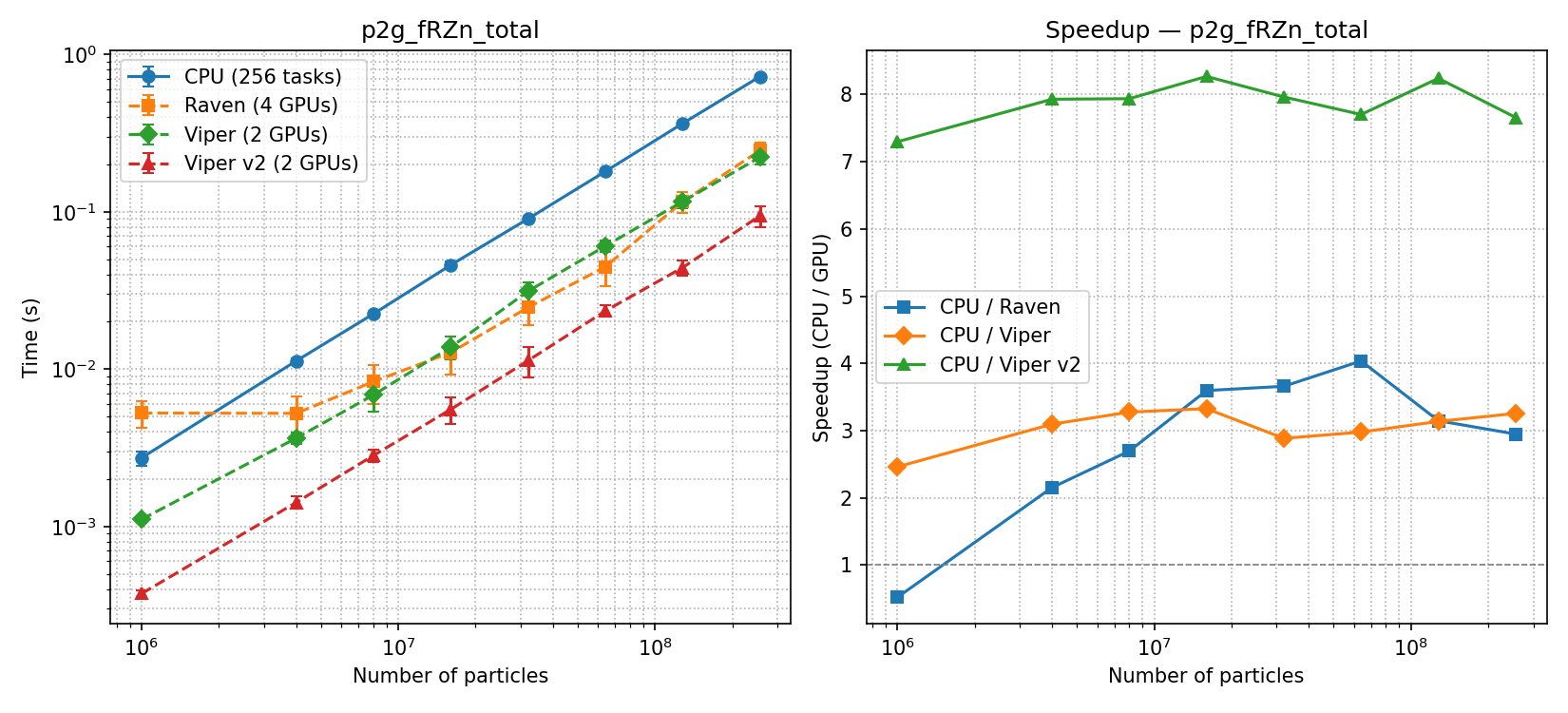}
  \caption{Left: wall-clock time of \texttt{p2g\_fRZn\_total} on TOK
    (256 MPI tasks), Raven (4 A100 GPUs), Viper (2 MI300A APUs,
    particle-parallel kernel) and Viper v2 (\texttt{collapse(3)} kernel).
    Right: speedup relative to the CPU baseline.}
  \label{fig:p2g_total_v2}
\end{figure}

Figure~\ref{fig:p2g_total_v2} shows the wall-clock time of the full \texttt{p2g\_fRZn} operation (left) and the corresponding speedup over the CPU baseline (right), as a function of the marker count $N_p$, for the two GPU kernel versions of Section~\ref{sec:nfa_offload} on Viper together with Raven and the TOK baseline. All three systems scale linearly with $N_p$ in the log--log plane, confirming that kernel execution time is proportional to the workload.

\begin{figure}[htbp]
  \centering
  \includegraphics[width=\textwidth]{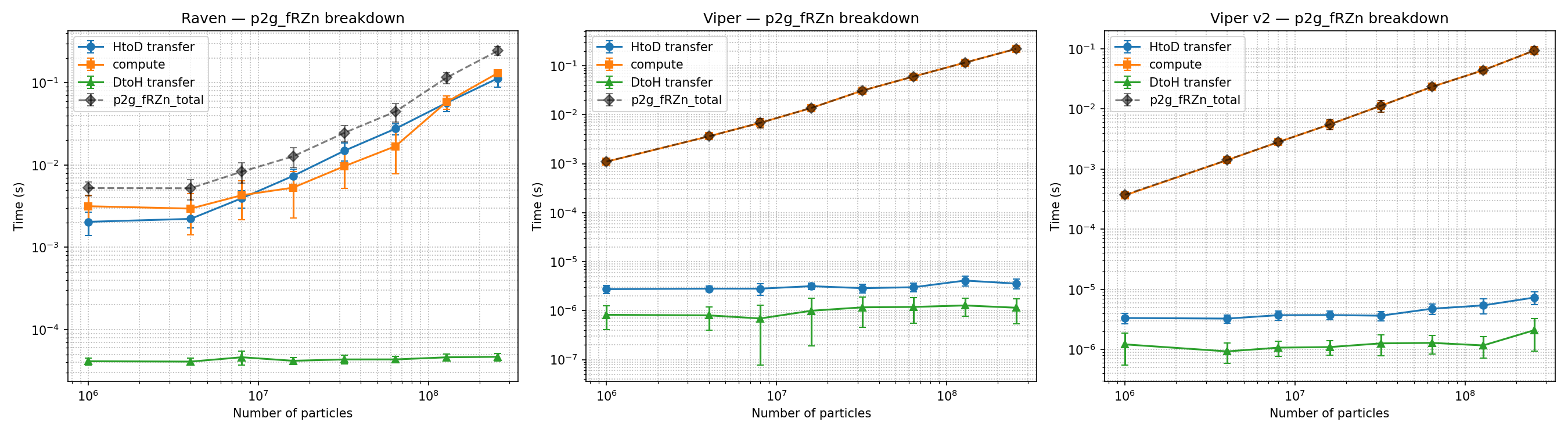}
  \caption{Phase breakdown of \texttt{p2g\_fRZn} as a function of the number of particles on Raven (left), Viper with the first kernel version (centre), and Viper with the \texttt{collapse(3)} redesign (v2, right): host-to-device (HtoD) transfer, kernel compute, device-to-host (DtoH) transfer, and total time. On Raven, the HtoD transfer over PCIe is comparable to the compute time at all sizes and therefore contributes substantially to the total. On Viper, the unified memory of the MI300A keeps both transfer phases flat at a few microseconds regardless of $N_p$, so the total time coincides with the compute time; the redesigned kernel (v2) lowers this compute time by $2.4$--$2.7\times$.}
  \label{fig:p2g_breakdown}
\end{figure}

The two GPU systems differ qualitatively at small marker counts because of their different memory architectures. On Raven, at $N_p=10^6$ the GPU run is slower than the CPU baseline (speedup $\approx 0.43$): the cost of transferring particle data to the discrete A100 over PCIe exceeds the compute savings at this size, and the break-even point lies between $4\times10^6$ and $8\times10^6$ markers; beyond it the speedup rises to a plateau of about $3.3\times$. On Viper the unified-memory MI300A pays essentially no transfer overhead even at $N_p=10^6$ (speedup already $\approx 2.1\times$), and the particle-parallel kernel plateaus around $3.4\times$. Figure~\ref{fig:p2g_breakdown} isolates the cause on Raven: the host-to-device transfer and the compute phase are comparable in magnitude throughout the scan, and at the largest size tested the transfer still accounts for roughly $46\%$ of the total time, whereas on Viper both transfer phases are flat at about $3\,\mu$s regardless of $N_p$, so the total time there is essentially the compute time.

The \texttt{collapse(3)} redesign of Section~\ref{sec:nfa_offload} noticeably improves kernel performance: it reduces the kernel time on Viper by $2.4$--$2.7\times$ across the marker-count scan (Figure~\ref{fig:p2g_total_v2}, Viper v2), raising the total speedup over the CPU baseline from about $3\times$ to $7$--$8\times$; at the largest size tested ($N_p=2.56\times10^8$) the Viper time drops from $0.222$\,s to $0.094$\,s, corresponding to a speedup over the CPU baseline of $\approx 7.7\times$. Correctness was verified element by element against the CPU reference at $N_p = 8\times10^6$: the projected density and the particle weights agree to a maximum relative error of $1.2\times10^{-11}$ and $1.1\times10^{-12}$ respectively, consistent with the different order of floating-point accumulation between the two architectures.

\subsection{Performance of particle projection: piecewise-field-aligned FEM}
\label{sec:perf_fa}

\begin{figure}[htbp]
  \centering
  \includegraphics[width=0.97\textwidth]{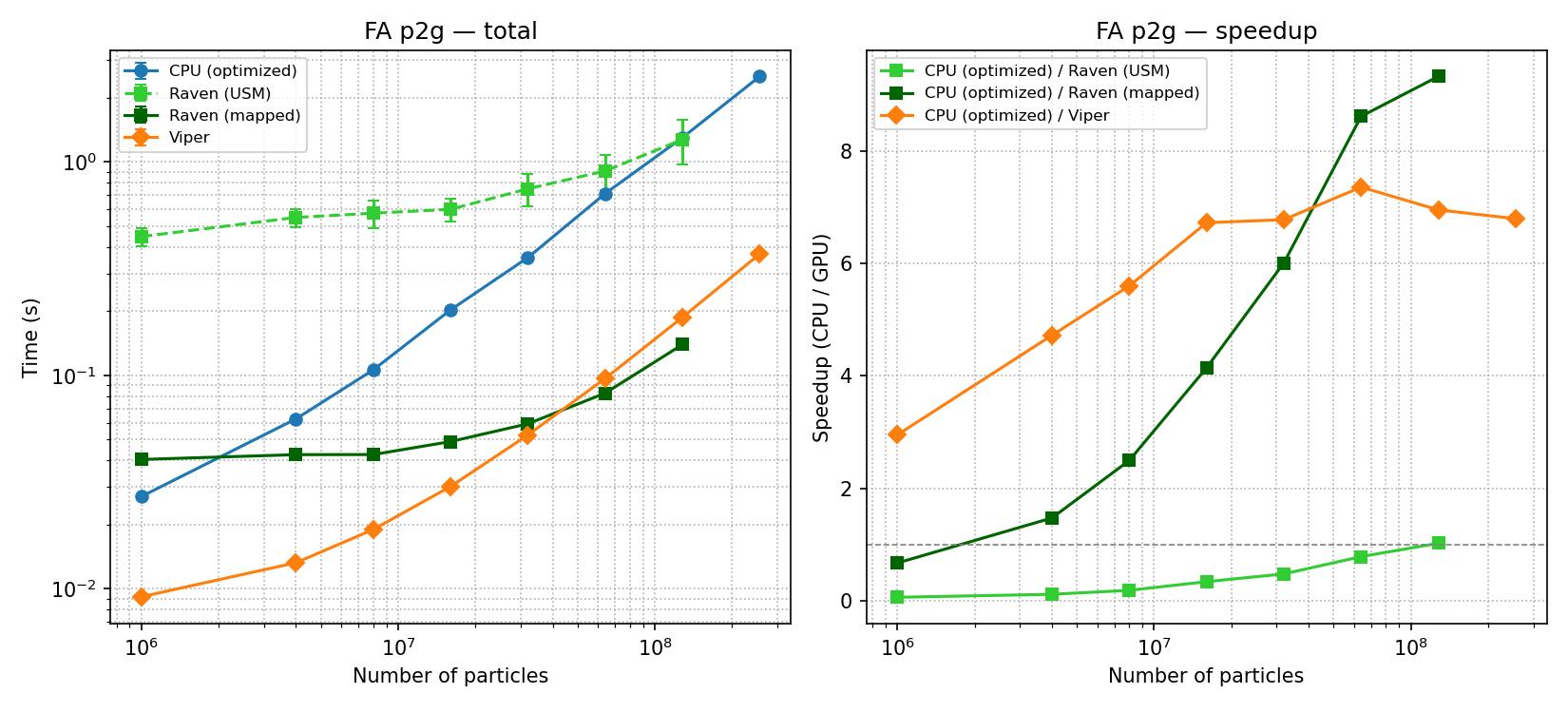}
  \caption{Total time per field-aligned p2g call (left) and speedup over the optimised CPU baseline (right, same spatial sort and B-spline evaluator as the GPU kernel), as a function of the marker count. GPU hardware: the A100 of Raven with managed memory and explicit mapping, and the MI300A of Viper with USM (Unified Shared Memory).}
  \label{fig:p2g_total_fa}
\end{figure}

The field-aligned projection is benchmarked on the same hardware, test case, and run parameters as the non-field-aligned kernel (Section~\ref{sec:testcase_hpc}), with the spatial sort of Section~\ref{sec:fa_offload} enabled and the marker count scanned from $1$ to $256$~million ($128$~million for the explicitly-mapped A100 runs). Because the A100 of Raven is a discrete GPU, two memory strategies are compared: managed memory, in which the driver migrates pages on demand and the unmodified APU-oriented code runs unchanged, and explicit mapping, in which the host--device transfers are issued by hand as for the NFA kernel. Figure~\ref{fig:p2g_total_fa} reports the total time per call (left) and the speedup over an \emph{optimised} CPU baseline that carries the same spatial sort and B-spline evaluator as the GPU kernel (right), which is the fairer comparison since a large part of the apparent gain over an unoptimised CPU baseline is due to the CPU-side optimisations of Section~\ref{sec:fa_offload} alone.

The A100 with managed memory is markedly the slowest GPU option, rising only from $\sim0.45$\,s at $1$\,M markers to $\sim1.3$\,s at $128$\,M, against $\sim0.14$\,s for the explicitly-mapped A100 at the same size --- a factor of $9$ to $13$ that is entirely the cost of page migration on otherwise identical kernels and launch configurations. Against the optimised CPU baseline, the explicit data map version on the A100 ranges from below unity at $1$\,M markers to $\sim9\times$ at $128$\,M, Viper ranges from $\sim3\times$ to a $\sim7\times$ plateau, and the managed memory A100 hovers around break-even throughout; two regimes are visible, with Viper the best device at small marker counts and the explicitly-mapped A100 overtaking it from a few tens of millions of markers upward. Once the CPU-side optimisations are accounted for, the actual hardware speedup of the field-aligned projection at production scale is therefore about $7$--$8\times$, matching the NFA kernel (Section~\ref{sec:perf_nfa}). Correctness was again verified element by element against the CPU reference ($8\times10^6$ markers, $483\,552$ degrees of freedom): out of $340\,656$ non-zero entries of the projected density, only $547$ show a relative deviation above $10^{-10}$.

\subsection{Performance of field solvers}
\label{sec:solver_results}

The benchmark compares one full node of three machines, the same
systems used for the projection kernels and described in
Section~\ref{sec:performance}. The problem size is scanned through
the number of toroidal planes, $n_\varphi = 8$ to $48$, i.e.\ from
$2.4\times10^5$ to $1.45\times10^6$ unknowns on the reference edge mesh
($6 N_{\mathrm{vert}} n_\varphi$ degrees of freedom with
$N_{\mathrm{vert}} = 5037$). $n_\varphi = 16$ is the production case of \cite{Lu2026}. Each
configuration performs five consecutive solves of the same system and the
reported time is the mean of solves 2--5 since the first solve pays one-time
costs (on the GPU, the host-to-device copy of the matrix) that do not
recur in a simulation, where the matrix stays resident.

On Viper, all results below use the Kokkos backend. The native HIP
backend was benchmarked on the same matrices. The Kokkos and HIP are close for a low and high number of toroidal planes (within $10\%$), but for intermediate values the native backend
degrades, up to $1.7\times$ slower at $n_\varphi = 24$ ($0.603$ vs
$0.356$~s). This finding supports the choice of Kokkos over HIP, highlighting PETSc's own MI300A development
targeting the former programming model (Section~\ref{sec:solver_petsc}).
On Raven, the native CUDA backend is used
(Section~\ref{sec:solver_petsc}).

Every run is
checked in two ways: the relative error of the solution against the
manufactured analytic solution, and the iteration count, which for a
given $n_\varphi$ and equal solver settings (method and preconditioner)
must agree across machines, backends, rank counts, and matrix formats up
to floating-point round-off. The validation shows that the counts agree to within
$\pm 2$ iterations at $n_\varphi = 12$--$20$, and to within $27$
iterations out of ${\sim}400$ ($7\%$) in the worst cases
($n_\varphi = 8$, $24$, $48$), where the residual stays close to the
tolerance over the last iterations and the stopping point becomes
sensitive to rounding.

\subsubsection{Choice of the Solver and preconditioner}
\label{sec:solver_results_choice}

The solvers and preconditioners addressed in this work are the following.
\begin{itemize}
    \item The Conjugate Gradient (CG) with Jacobi preconditioner.
    \item The GMRES with Jacobi/block-Jacobi/ILU(0) preconditioner.
\end{itemize} 

The solver initially used in this work was CG with Jacobi
preconditioning, on the assumption that the operator is symmetric. CG
converges at most sizes, but at $n_\varphi = 24$ it
stagnates at the iteration limit without converging, identically on all
three machines and with independently assembled matrices. The explanation
is that the field-aligned operator is not symmetric: the relative
asymmetry in the Frobenius norm,
$\|\mathsf{L}-\mathsf{L}^{T}\|_F / \|\mathsf{L}\|_F$, is
$\approx 1.6\%$ at every $n_\varphi$, so the necessary condition to apply CG is violated. In most cases the asymmetry happens to be harmless, while at
$n_\varphi = 24$ it is fatal.
All results below therefore use GMRES, which requires no symmetry and
converges at every size tested (378--552 iterations at
$\mathrm{rtol}=10^{-5}$).

On the CPU node the natural reference is PETSc's parallel default,
GMRES with block-Jacobi/ILU(0). It is
the fastest CPU option where it works, but the ILU(0)
factorisation hits a zero pivot and fails for $n_\varphi \ge 20$. The GMRES+Jacobi configuration achieves convergence at every size. The CPU reference below is
therefore the default where it converges ($n_\varphi \le 16$) and
GMRES+Jacobi elsewhere. On the GPUs the preconditioner is Jacobi in all
cases, since the incomplete factorisations have no device implementation
(Section~\ref{sec:solver_petsc}). Following these findings, the
production runs of the following sections set GMRES+Jacobi explicitly on
every architecture.

\subsubsection{The Number of MPI Ranks}
\label{sec:solver_results_ranks}

The GPU runs expose one tuning parameter with a surprisingly large
effect: the number of MPI ranks sharing the GPUs. Thanks to the
layout-independent matrix format
(Section~\ref{sec:solver_petsc}), the same stored matrix can be
solved with any rank count, so this parameter could be scanned directly.
On the Viper node, one rank per APU, for a total of 2, is the best choice at
every size; doubling to 4 ranks already incurs a slowdown factor $1.3$ to
$2.1$. The Raven node shows exactly the same behaviour (1 rank per GPU is
best, 2 per GPU is $1.5$--$3.1\times$ slower). The reason is the structure of
GMRES: each iteration performs several global reductions
(orthogonalisation, norms) whose latency grows with the number of ranks,
while the sparse matrix--vector product saturates the GPU bandwidth
already with one rank per device. Since the particle stages of the code require several ranks per node and the solver only one per GPU device, the solution is a dedicated communicator for the latter (a subset of ranks performing the solve)
as its natural integration into fully GPU-offloaded production runs.

\subsubsection{Performance Comparison of CPU and GPU Nodes}
\label{sec:solver_results_nodes}

\begin{figure}[htbp]
  \centering
  \includegraphics[width=0.85\textwidth]{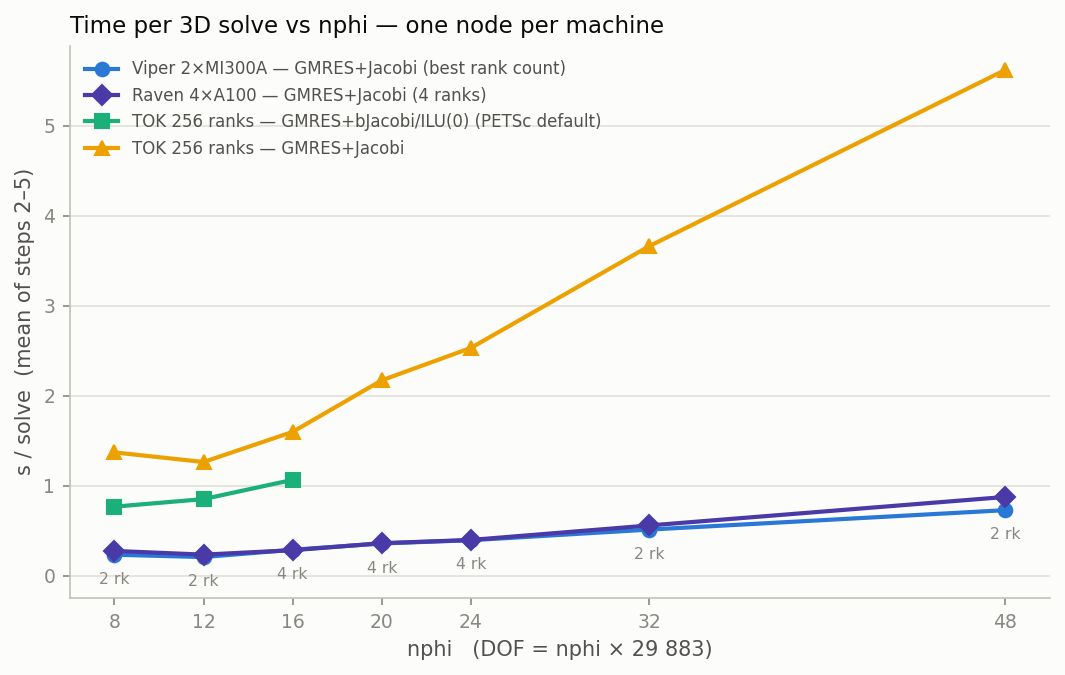}
  \caption{Time per 3D solve versus the number of toroidal planes
  $n_\varphi$, one full node per machine. GPU points use the best rank
  count (2 ranks everywhere for Viper, 4 for Raven). The PETSc
  default on TOK (green line) stops at $n_\varphi = 16$, where its ILU(0)
  factorisation starts to fail.}
  \label{fig:solver_times}
\end{figure}

\begin{figure}[htbp]
  \centering
  \includegraphics[width=0.85\textwidth]{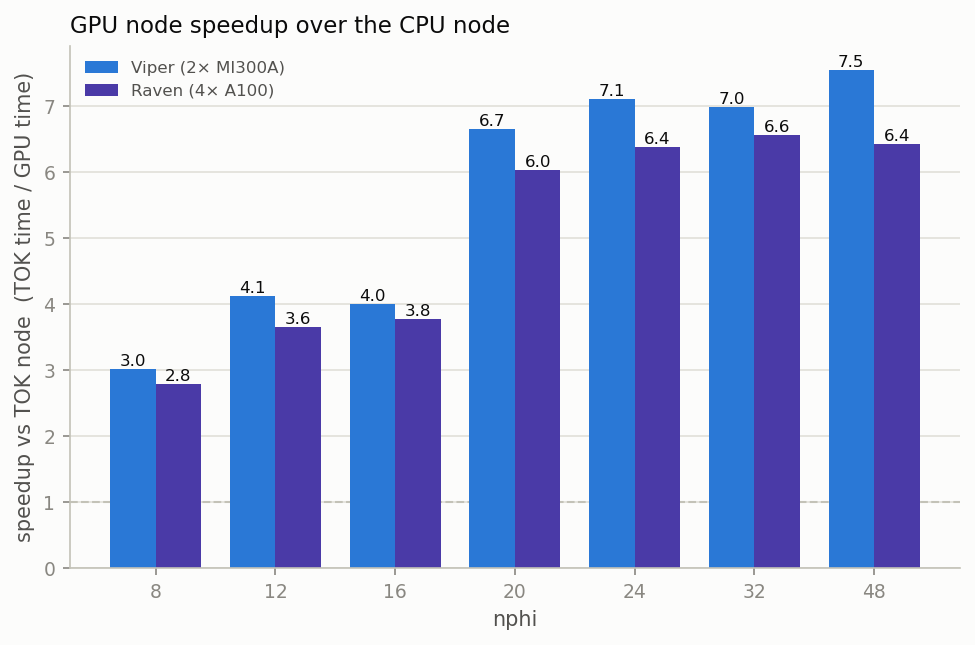}
  \caption{Speedup of one GPU node (Viper: 2$\times$ MI300A; Raven:
  4$\times$ A100) over one TOK CPU node (256 cores), best configuration
  on each side.}
  \label{fig:solver_speedup}
\end{figure}

Figure~\ref{fig:solver_times} shows the time per solve of the best
configuration of each node, while Figure~\ref{fig:solver_speedup} presents the
corresponding speedup of the two GPU nodes over the CPU node. 
The corresponding numerical values are given in Table~\ref{tab:solver_results}.

\begin{table}[h]
\centering
\caption{Time per solve (s, mean of solves 2--5) for the best
configuration of each node, and speedup of the GPU nodes over the CPU
node. In parentheses, the preconditioner
(TOK; J = Jacobi, the PETSc default fails for $n_\varphi \ge 20$). The best rank count is always 2 on a Viper node, and 4 on a Raven node, matching the number of available devices.}
\vspace{5pt}
\label{tab:solver_results}
\begin{tabular}{r r r r r r r}
\toprule
$n_\varphi$ & DOF & Viper & Raven & TOK & \multicolumn{2}{c}{speedup} \\
 & & & & & Viper & Raven \\
\midrule
 8 & $241\,776$   & 0.254 & 0.274 & 0.766 (ILU) & 3.0 & 2.8 \\
12 & $362\,664$   & 0.206 & 0.233 & 0.85\phantom{0} (ILU) & 4.1 & 3.6 \\
16 & $483\,552$   & 0.266 & 0.282 & 1.064 (ILU) & 4.0 & 3.8 \\
20 & $604\,440$   & 0.326 & 0.360 & 2.17\phantom{0} (J) & 6.7 & 6.0 \\
24 & $725\,328$   & 0.356 & 0.396 & 2.53\phantom{0} (J) & 7.1 & 6.4 \\
32 & $967\,104$   & 0.524 & 0.557 & 3.66\phantom{0} (J) & 7.0 & 6.6 \\
48 & $1\,450\,656$ & 0.745 & 0.874 & 5.62\phantom{0} (J) & 7.5 & 6.4 \\
\bottomrule
\end{tabular}
\end{table}

\paragraph{Scaling with the problem size.}
The CPU time increases  approximately linearly with the number of unknowns, as expected for
a bandwidth-bound iteration on a node that is already saturated. In contrast, the GPU nodes exhibit substantially weaker growth. Increasing the problem size by a factor of six, from $n_\varphi=8$ to $48$, increases the solve time on
Viper by only a factor of about three, indicating that the smaller
problems do not fully utilise the available GPU resources. The speedup
therefore grows with the problem size, from $\approx 3\times$ at the
smallest cases to $7.5\times$ (Viper) and $6.4\times$ (Raven) at the
largest, Raven peaking at $6.6\times$ for $n_\varphi = 32$. The trend is
not strictly monotonic at the low end: both GPU nodes are slower at
$n_\varphi = 8$ than at $n_\varphi = 12$ ($0.254$ vs $0.206$~s on Viper,
$0.274$ vs $0.233$~s on Raven). This is the same underfilling seen from
the other side: at the smallest size the solve is dominated by the
per-iteration latency floor --- kernel launches and the global reductions
of the Krylov iteration --- which does not shrink with the problem, so
the measured time follows the iteration count rather than the work per
iteration.

\paragraph{Equivalent performance for Viper and Raven.}
At equal solver settings, the performance on Viper and Raven nodes is very similar across the
entire scan, with Viper consistently faster by $6$--$17\%$ (e.g.\ $0.266$
vs $0.282$~s at $n_\varphi = 16$, $0.356$ vs $0.396$~s at $24$). Thus, for this workload, one node
with two MI300A matches one node with four A100 on this workload. 

\paragraph{PETSc binary I/O substantially reduces memory usage.}
The new restart mechanism also provides a significant reduction in memory
consumption, which is crucial for the largest case. With the original per-rank restart files, the $n_\varphi = 48$ solve did not fit on a single Viper node:
of the $2\times128$~GB of HBM3 installed per node, the batch system
makes at most $220$~GB allocatable to a job \cite{viper_docs}, the rest
being reserved for the system. The matrices are loaded into a structure preallocated with a
large upper-bound estimate of the nonzeros per row ($\approx 19\times$
the actual count), inflating the footprint to $\approx 29$~GB per rank.
Loading the same matrix through \texttt{MatLoad} from the PETSc binary
format allocates exactly the stored nonzeros: the identical solve then
takes $6.3$~GB per rank, a $\approx 4.6\times$ reduction, and the full
node-versus-node scan could be completed on one node. In a production
context, the same observation applies to the assembly preallocation and is
an easy target for future memory savings.

\subsection{Full-simulation performance}
\label{sec:fullsim_performance}

\begin{figure}[htbp]
  \centering
  \includegraphics[width=\textwidth]{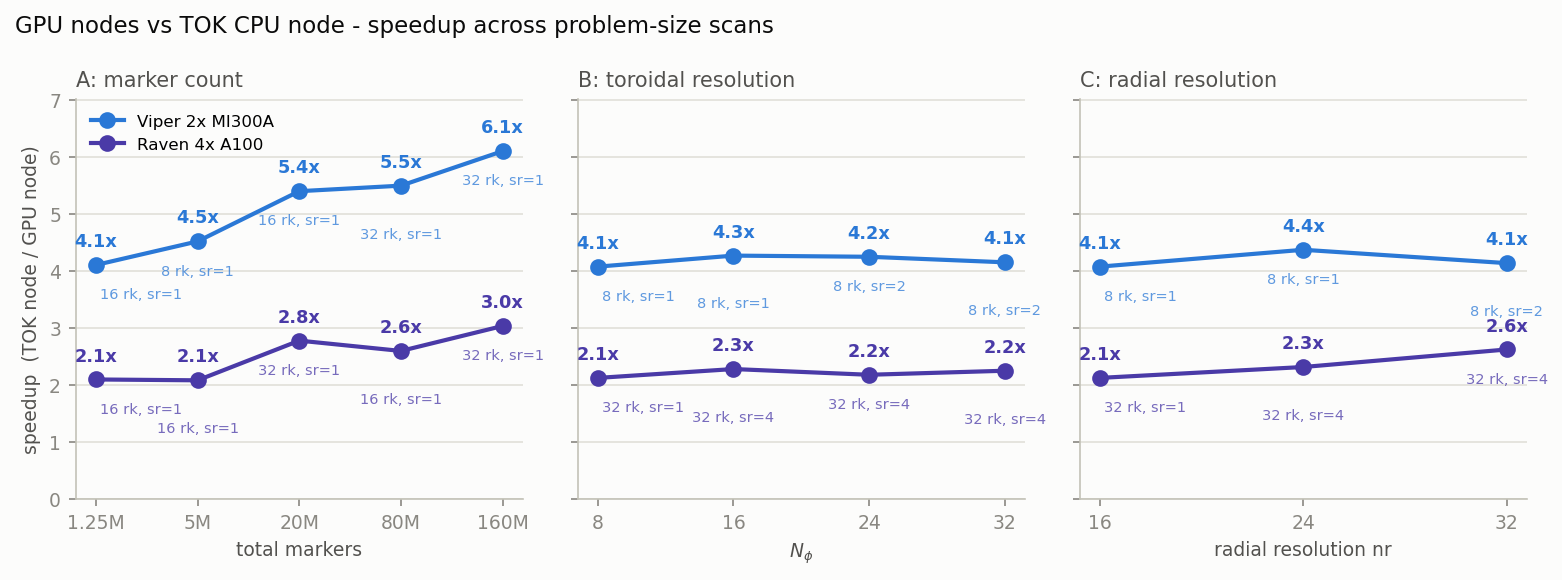}
  \caption{Full simulation speedup of one GPU node over one TOK CPU
    node (256 ranks), for the best configuration of each point. The
    labels below the points give that configuration (MPI ranks and
    solver ranks). A: marker-count scan on the base grid. B, C:
    toroidal and radial resolution scans at 10 million markers.}
  \label{fig:fullsim_speedup_scans}
\end{figure}

The individual kernels of Sections~\ref{sec:perf_nfa}--\ref{sec:solver_results} are combined into the fully GPU-resident time loop of Section~\ref{sec:fullsim} and benchmarked on the complete TCV-X21 case, using the solver sub-communicator and the Raven data-management scheme of Section~\ref{sec:fullsim} with the problem-size-dependent rank counts that give the best performance on each machine. Figure~\ref{fig:fullsim_speedup_scans} collects the node-versus-node speedup of one GPU node over one $256$-core TOK node: the marker count grows from $1.25$ to $160$ million on the base grid (panel A), and the toroidal and radial resolution are scanned separately at $10$ million markers (panels B, C). As with the stand-alone solver of Section~\ref{sec:solver_results}, the speedup grows with the problem size on both machines, from $4.1\times$ to $6.1\times$ on Viper and from $2.1\times$ to $3.0\times$ on Raven, because the smaller cases do not fill the GPUs; across the resolution scans, with the solver sub-communicator tuned for each grid, the speedup stays close to $4.1$--$4.4\times$ on Viper and $2.1$--$2.6\times$ on Raven.

\begin{figure}[htbp]
  \centering
  \includegraphics[width=\textwidth]{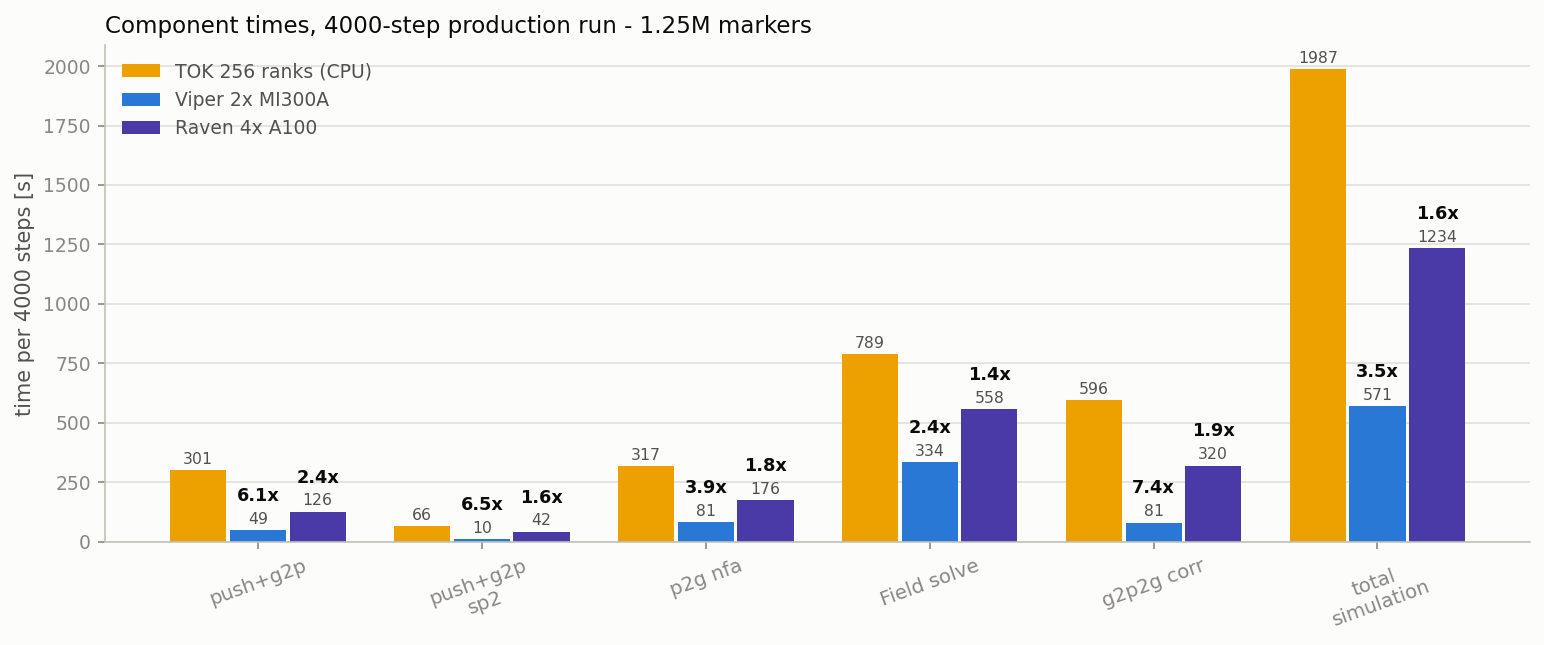}
  \caption{Component times of the 4000-step production run
    ($1.25$ million markers) on the three machines (TOK 256 ranks,
    Viper 8, Raven 16). Bold labels give the speedup of each component
    over TOK. The field solve dominates the GPU runs at this problem
    size.}
  \label{fig:fullsim_kernels_prod4k}
\end{figure}

\begin{figure}[htbp]
  \centering
  \includegraphics[width=\textwidth]{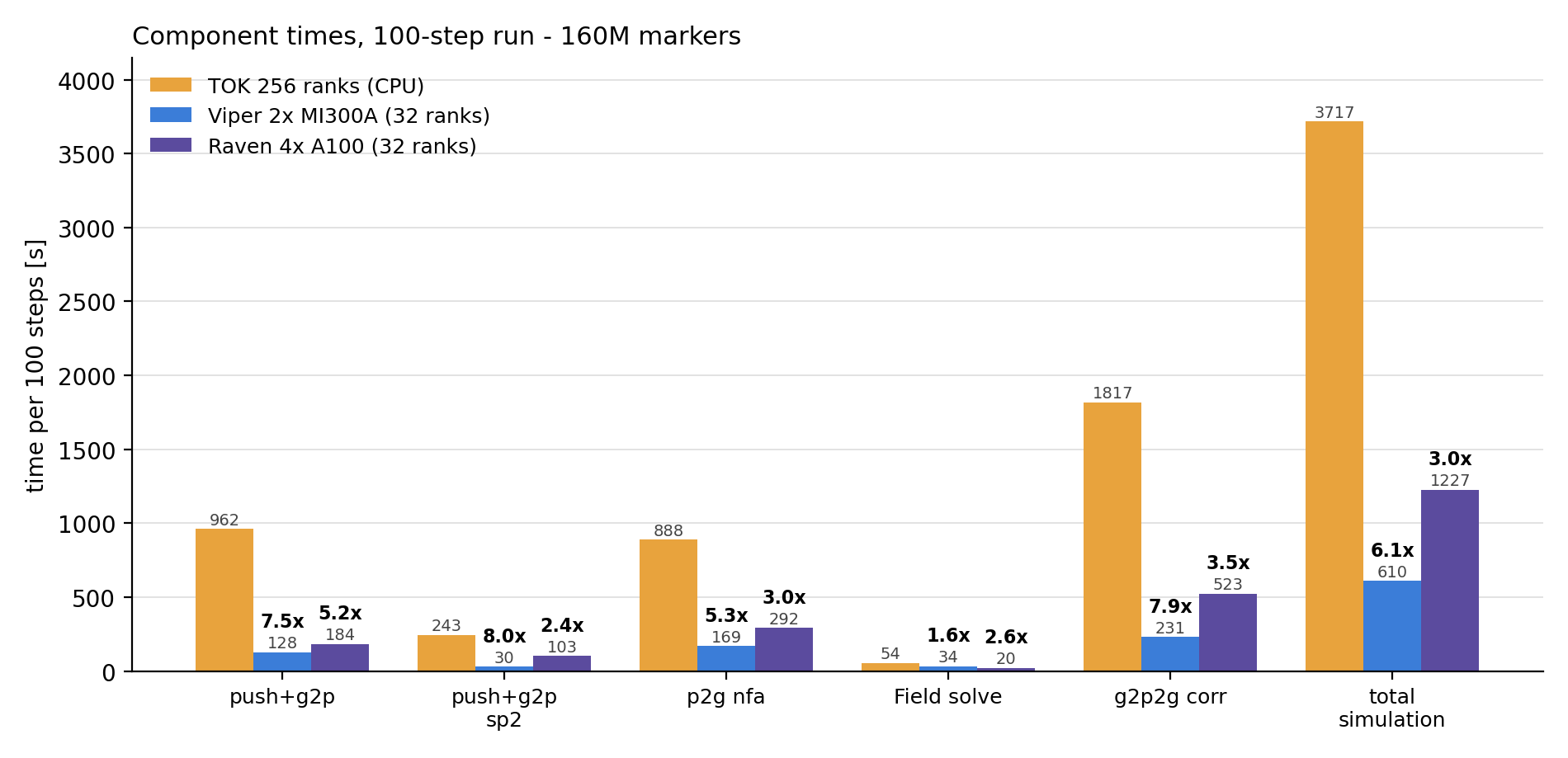}
  \caption{Component times of a $100$-step segment of the
    $160$-million-marker case on the three machines (TOK 256 ranks,
    Viper 32, Raven 32). Bold labels give the speedup of each component
    over TOK. At this marker count the particle stages dominate the CPU
    run and the field solve becomes negligible.}
  \label{fig:fullsim_kernels_160M}
\end{figure}

Figure~\ref{fig:fullsim_kernels_prod4k} breaks the $4000$-step production run used for the validation of Section~\ref{sec:sim_results} down into its main components. The whole run takes $1987$\,s on TOK, $571$\,s on Viper ($3.5\times$) and $1234$\,s on Raven ($1.6\times$); the field solve is the largest single component on all three machines at this marker count ($789$, $334$ and $558$\,s) and the least accelerated ($2.4\times$ on Viper, $1.4\times$ on Raven), consistent with the latency floor identified for the stand-alone solver in Section~\ref{sec:solver_results}. Figure~\ref{fig:fullsim_kernels_160M} repeats the breakdown at $160$ million markers, where the picture changes significantly: the particle stages now dominate the CPU time and are precisely where the GPUs excel, so the node-level speedup approximately doubles, to $6.1\times$ on Viper and $3.0\times$ on Raven, while the field solve amounts to a small fraction of the time step.

The practical consequence of these two regimes is substantial: extrapolating the per-step cost of the $160$-million-marker case, a $4000$-step run at that scale would take roughly $41$ hours on a TOK node against about $7$ hours on one Viper node and $14$ on Raven, turning a marker count that is impractical on a CPU node into a routine overnight run.

\section{Simulation results of instabilities in tokamaks}
\label{sec:sim_results}

The performance results of Section~\ref{sec:performance} do not by
themselves establish that the fully integrated simulation is correct. The
port is therefore validated on the physical observables of an
ion-temperature-gradient (ITG) instability in the TCV-X21 configuration,
without and with the open-field-line region, comparing the TOK CPU node
with the Viper (AMD MI300A) and Raven (NVIDIA A100) GPU nodes.

The markers are loaded from per-marker random streams addressed by the
global marker index: runs on different machines and rank counts start from
the same initial state, and their energy traces can be compared point by
point. The most sensitive observable is the growth rate $\gamma_P$, half
the slope of $\ln E$ in the linear phase, to which every term of the
gyrocentre equations of motion contributes at every substep.

\subsection{Core plasma simulation without open field lines}

\begin{figure}[htbp]
  \centering
  \includegraphics[width=0.6\textwidth]{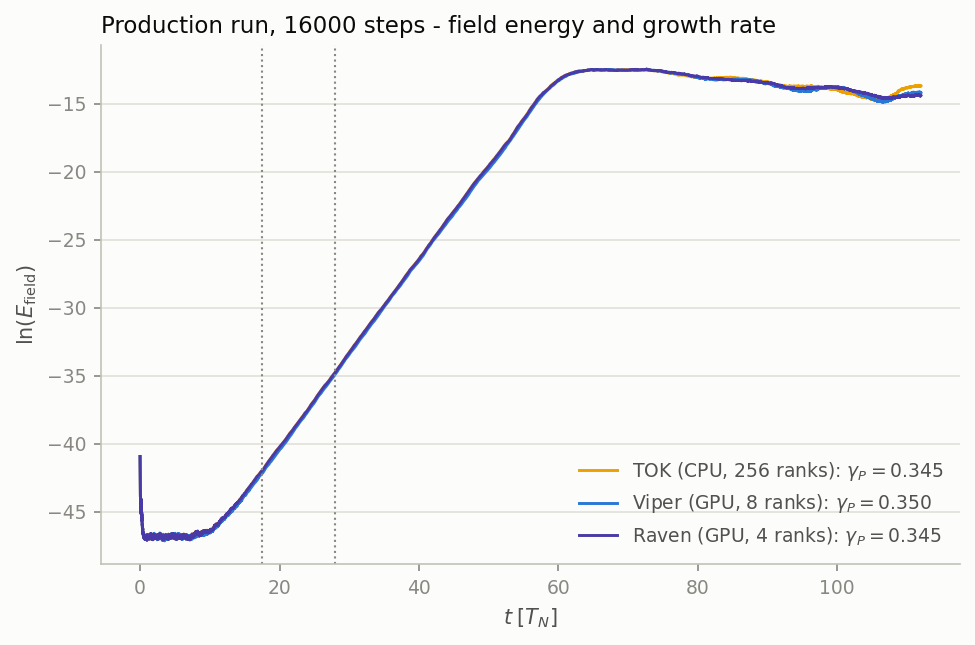}
  \includegraphics[width=0.8\textwidth]{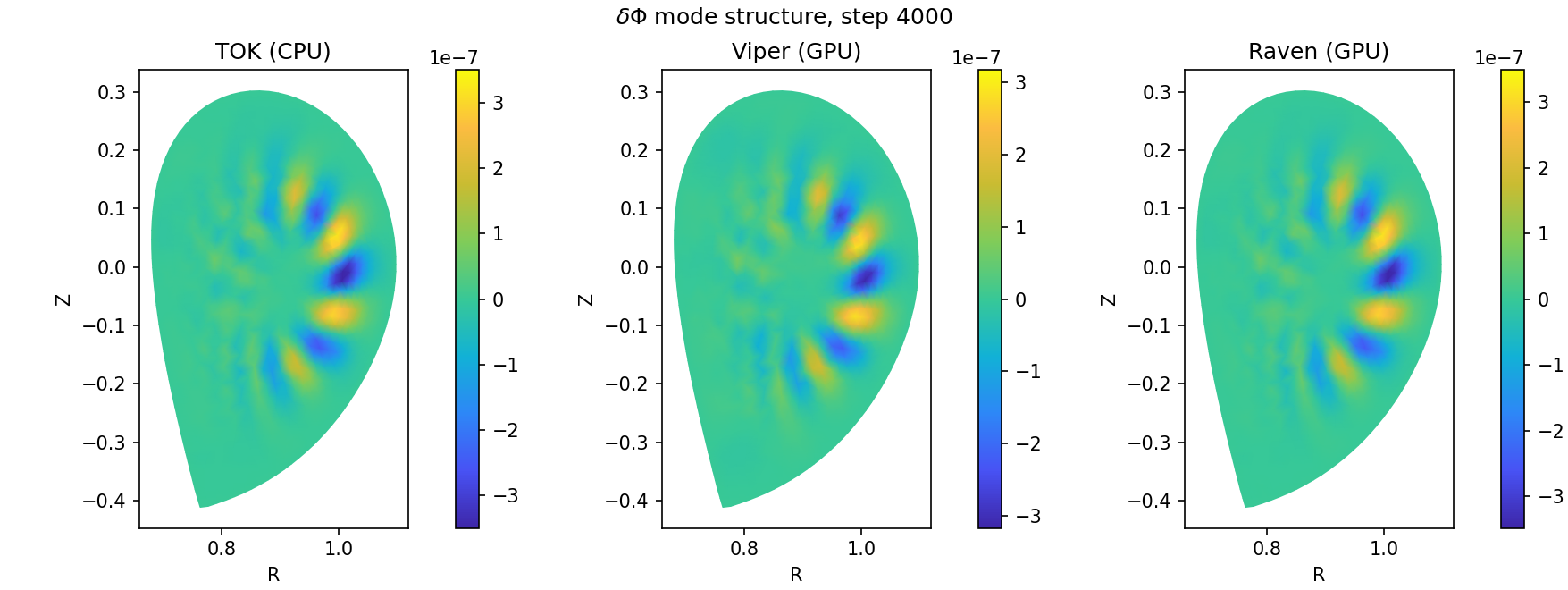}
  \caption{Upper: field energy of the case without open field lines ($1.25$
    million markers, $\Delta t = 0.007\,T_N$, $16000$ steps) on the three
    machines. Dotted lines mark the fit window ($17.5$--$28\,T_N$) of the
    growth rates in the legend. The TOK and Raven curves overlap.
    Lower: poloidal structure of $\delta\Phi$ at step $4000$ (linear
    phase) of the case without open field lines (first toroidal plane,
    vertex values).}
  \label{fig:core_energy}
\end{figure}

The electrostatic case uses the $1.25$ million markers of
Section~\ref{sec:fullsim_performance}, run for $16000$ steps of
$\Delta t = 0.007\,T_N$. The field energy (upper frame of Figure~\ref{fig:core_energy})
shows the decay of the initial perturbation, the noise floor of the marker
sampling, the exponential growth of the ITG mode and its nonlinear
saturation at $t \approx 60\,T_N$. The three machines give the same growth
rate to within $1.5\%$ ($\gamma_P = 0.345$, $0.350$ and $0.345$ on TOK,
Viper and Raven) and saturate at the same level. The TOK and Raven traces
coincide to within $0.03$ in $\ln E$ over the whole linear phase, while the
Viper trace deviates by up to $0.3$ in the noise floor. All three machines
produce the same ballooning ITG mode (lower frame of Figure~\ref{fig:core_energy}), with
the same poloidal mode number and localisation.

\subsection{Whole volume simulation with open field lines}

\begin{figure}[htbp]
  \centering
  \includegraphics[width=0.6\textwidth]{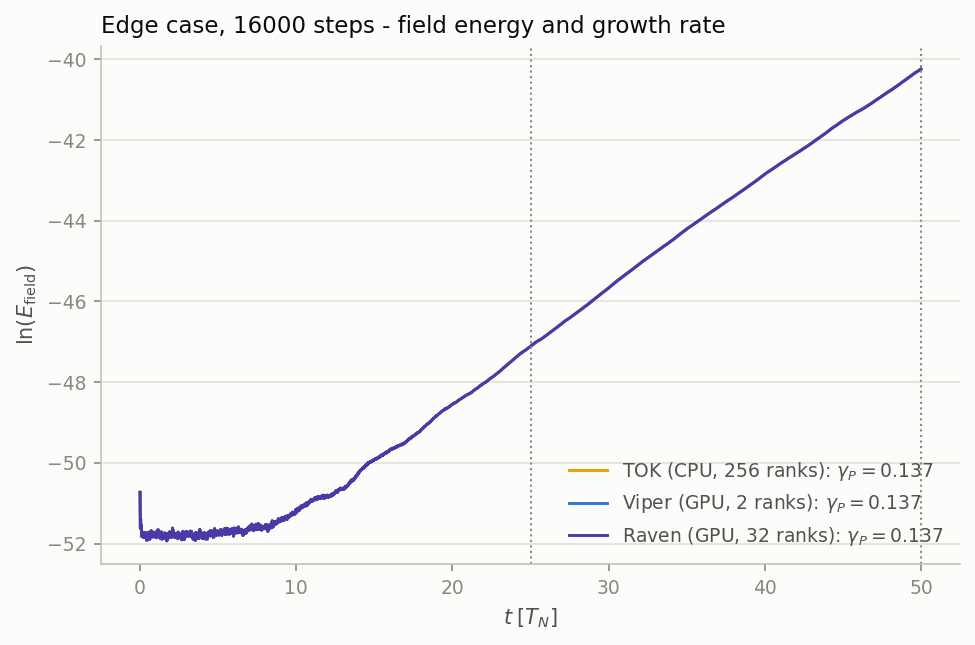}\\[2mm]
  \includegraphics[width=.8\textwidth]{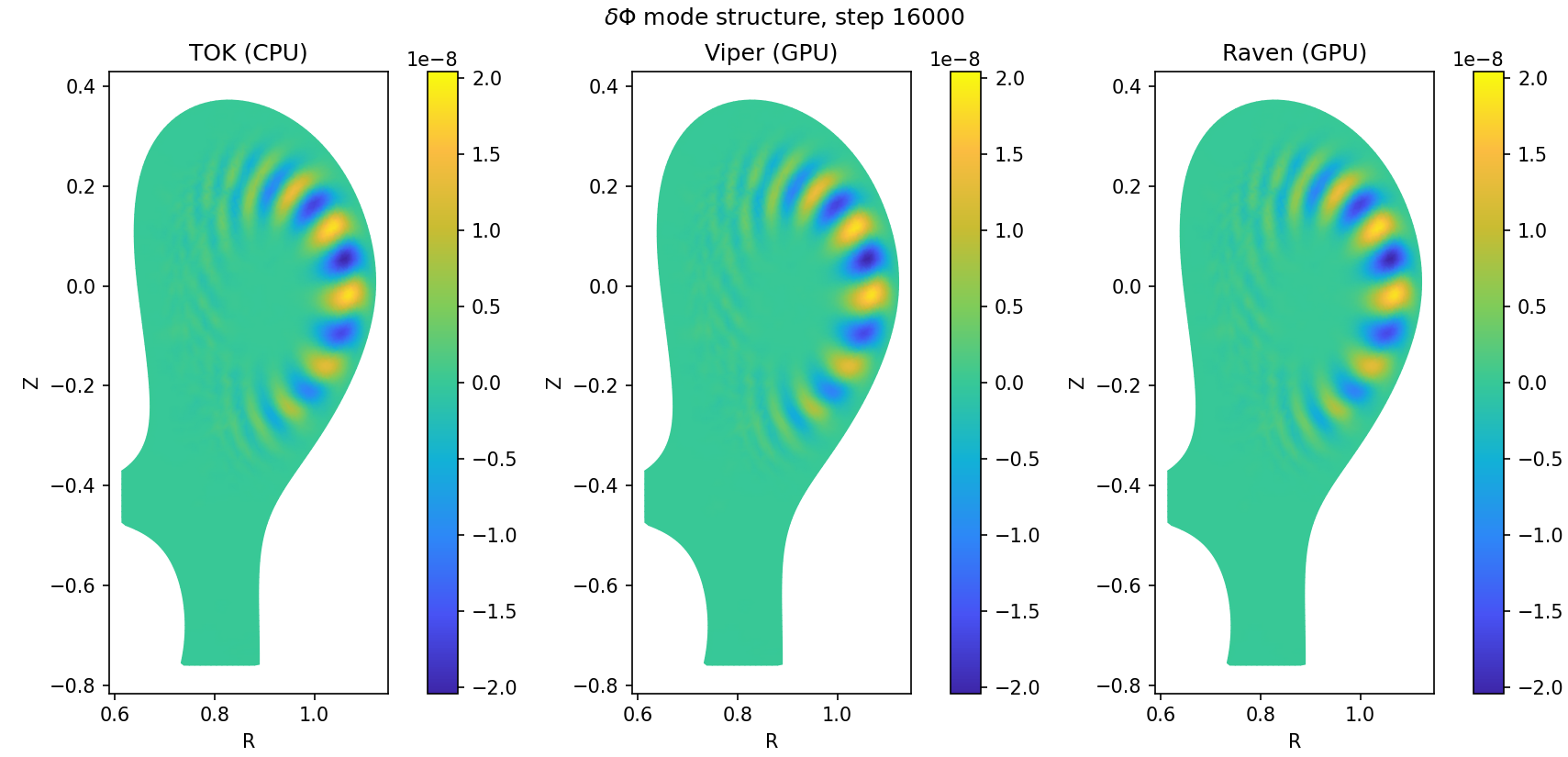}\\[2mm]
  \includegraphics[width=.8\textwidth]{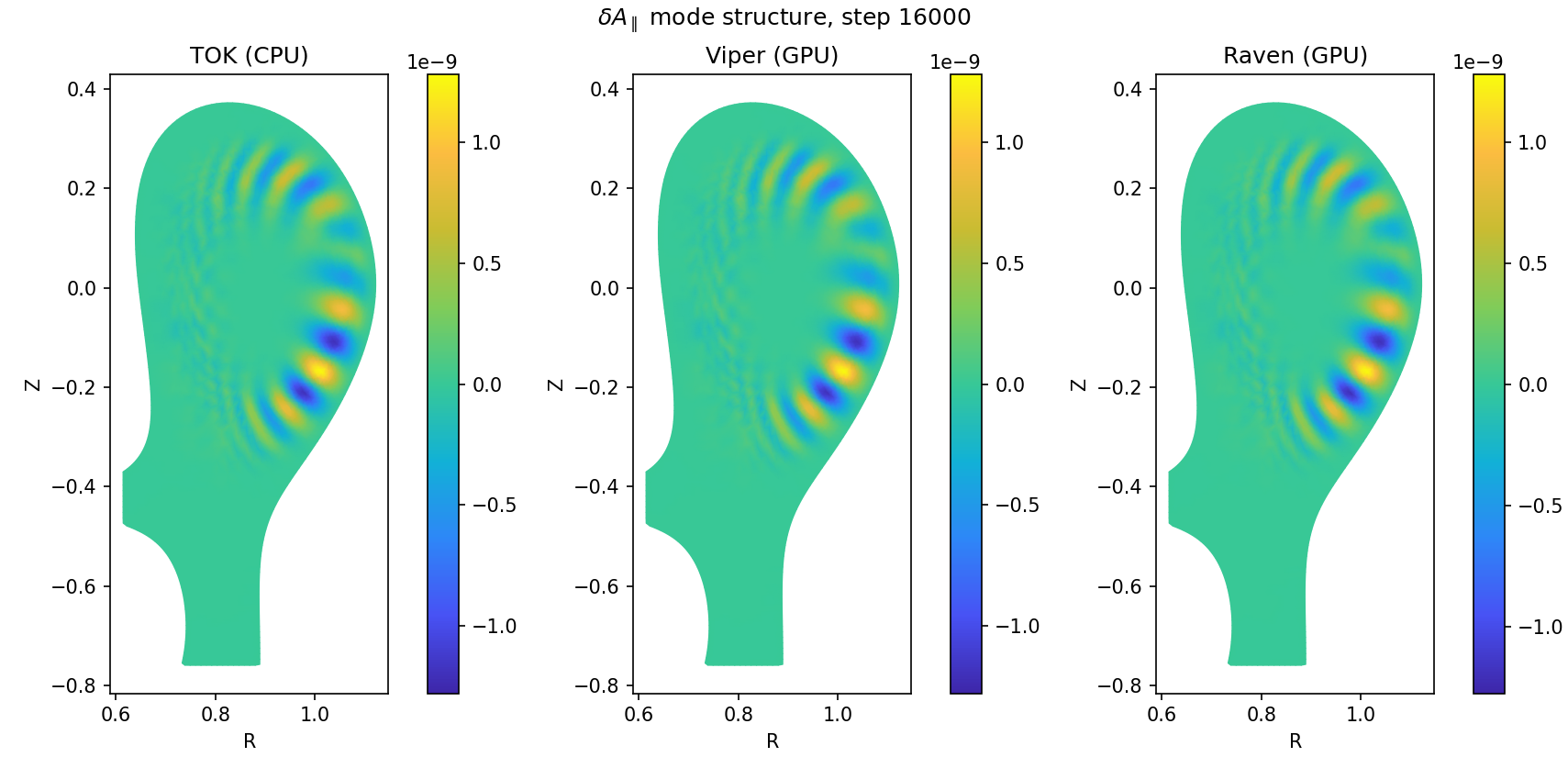}
  \caption{Upper: field energy of the electromagnetic case with open field
    lines ($20$ million markers, $\Delta t = 0.003125\,T_N$, $16000$ steps).
    The three curves overlap to within the line width, so only the last
    one drawn (Raven) is visible. Dotted lines mark the fit window
    ($25$--$50\,T_N$).
    Middle and lower: poloidal structure of $\delta\Phi$ (middle) and
    $\delta A_\parallel$ (lower) at step $16000$ of the case with open
    field lines (first toroidal plane, vertex values).}
  \label{fig:edge_energy}
  \label{fig:edge_modes}
\end{figure}

The second case adds the X point and the open-field-line region, is
electromagnetic, and uses $16$ million electron and $4$ million ion
markers, run for $16000$ steps of $\Delta t = 0.003125\,T_N$. Here the
three machines agree point by point: the energy traces coincide to within
$0.002$ in $\ln E$ over all $16000$ steps, noise floor included
(Figure~\ref{fig:edge_energy}), the growth rate is $\gamma_P = 0.137$ on
all three, and the structures of both $\delta\Phi$ and
$\delta A_\parallel$ are indistinguishable (Figure~\ref{fig:edge_modes}).

Together, these results establish that the GPU port reproduces the physics
of ion temperature gradient mode using TRIMEG-C1, in both the electrostatic and the electromagnetic regime, to
the accuracy needed for production instability studies.
 
\section{Summary and Outlook}
In this paper, the whole-device gyrokinetic code TRIMEG-C1 was rewritten in C++ as TRIMEG++ and ported to GPUs using OpenMP target offloading throughout, so that a single code base runs on both the unified-memory AMD MI300A APUs of Viper and the discrete NVIDIA A100 GPUs of Raven. In a gyrokinetic particle-in-cell simulation, each time step deposits the charge and current of the markers on the mesh, solves the field equations on that mesh, and advances the markers in the resulting fields; all three stages now run on the GPU. All speedups are reported as node-to-node comparisons, i.e.\ one GPU node (two MI300A on Viper, four A100 on Raven) against one full 256-core CPU node of the TOK cluster.

The particle-to-grid projection, in both its non-field-aligned and field-aligned formulations, reaches $7$--$8\times$ and $7$--$9\times$ respectively over an equally optimised CPU baseline at production marker counts. Two of the optimisations, a native B-spline evaluator and a spatial sort of the markers, also benefit the CPU code. The PETSc-based field solver runs on the GPU through the library's runtime-selectable backends and solves the field systems $3$ to $7.5\times$ faster than the CPU node in the stand-alone benchmark. Inside the integrated time loop, however, the mesh of the production case is too small to keep the GPU busy, so the solve is limited by fixed per-iteration costs rather than by computation, and its gain falls to $2.4\times$ on Viper and $1.4\times$ on Raven. The solve was also found to be fastest with one MPI rank per GPU, whereas the particle kernels benefit from many ranks; a dedicated solver sub-communicator resolves this conflict. With the particle pusher integrated into the same GPU-resident loop, the complete simulation runs $4.1$--$6.1\times$ faster on Viper and $2.1$--$3.0\times$ faster on Raven than the CPU node, with the speedup growing with problem size: a run with 160 million markers that would take about 41 hours on a CPU node completes in about 7 hours on one Viper node. On the hardware side, the shared CPU--GPU memory of the MI300A removes data transfers entirely, while the A100 reaches comparable performance only when transfers are managed explicitly. Over 16000-step TCV-X21 simulations, the GPU implementation reproduces the CPU physics, with growth rates agreeing within $1.5\%$ and identical energy evolution and mode structure.

Future optimisation will focus on the field solver, to enhance the code capability for simulations of large tokamak devices such as ITER and DEMO. At moderate marker counts the field solver is the largest single component of the time step and the least accelerated, since its cost is set by the mesh rather than by the number of markers, and only the Jacobi preconditioner currently has a GPU implementation in PETSc. All benchmarks in this paper are single-node, so multi-node scaling of the offloaded kernels remains to be characterised, and the optimal rank and solver-rank counts reported here are specific to the TCV-X21 case and the grid resolutions tested. Keeping the particle and field data resident on the device for the whole run, rather than remapping them at every kernel launch, is the most promising remaining optimisation, particularly for the small and medium-sized cases on Raven, and is the natural next step now that correctness and single-node performance have been established.

\section*{Conflict of Interest Statement}

The authors declare that the research was conducted in the absence of any commercial or financial relationships that could be construed as a potential conflict of interest.

\section*{Author Contributions}

N. Noventa: Software, Formal analysis, Investigation, Data Curation, Writing – Original Draft\\
G. Daneri:  Software, Formal analysis, Investigation, Data Curation, Writing – Original Draft\\
Z. X. Lu: Conceptualization, Software, Supervision, Writing – Original Draft, Writing – review \& editing\\
M. Hoelzl: Writing – Review \& Editing, Resources



\section*{Acknowledgments}

The simulations in this work were run on the local TOK cluster and the MPCDF Viper/Raven supercomputers.
We acknowledge the contribution of Johanna Potyka from AMD concerning the optimisation of the non-field-aligned deposition kernel on the MI300A platform. We further acknowledge her help in debugging the pusher kernel behavior on the same architecture, which eventually lead to a bug fix.
This work has been carried out within the framework of the EUROfusion Consortium, funded by the European Union via the Euratom Research and Training Programme (Grant Agreement No 101052200—EUROfusion). Views and opinions expressed are, however, those of the author(s) only and do not necessarily reflect those of the European Union or the European Commission. Neither the European Union nor the European Commission can be held responsible for them.

Figure~\ref{fig:PIC_loop} was generated with ChatGPT (GPT-5.5, Instant model, free web version; OpenAI) on 24 July 2026; the authors subsequently corrected it manually to ensure that it accurately represents the PIC cycle described in the text.

\section*{Data Availability Statement}
The data and code supporting the findings of this study are available from the corresponding author upon reasonable request.

\bibliographystyle{harvard-preprint}
\bibliography{reference}

\begin{thebibliography}{32}
\providecommand{\natexlab}[1]{#1}
\expandafter\ifx\csname urlstyle\endcsname\relax
  \providecommand{\doi}[1]{doi:\discretionary{}{}{}#1}\else
  \providecommand{\doi}{doi:\discretionary{}{}{}\begingroup
  \urlstyle{rm}\Url}\fi
\providecommand{\selectlanguage}[1]{\relax}
\providecommand{\bibAnnoteFile}[1]{%
  \IfFileExists{#1}{\begin{quotation}\noindent\textsc{Key:} #1\\
  \textsc{Annotation:}\ \input{#1}\end{quotation}}{}}
\providecommand{\bibAnnote}[2]{%
  \begin{quotation}\noindent\textsc{Key:} #1\\
  \textsc{Annotation:}\ #2\end{quotation}}

\bibitem[{{Advanced Micro Devices, Inc.}(2025)}]{amd_mi300a_datasheet}
[Dataset] {Advanced Micro Devices, Inc.} (2025).
\newblock {AMD Instinct MI300A APU Data Sheet}.
\newblock GD-83/GD-176
\bibAnnoteFile{amd_mi300a_datasheet}

\bibitem[{Balay et~al.(2025)Balay, Abhyankar, Adams, Benson, Brown, Brune
  et~al.}]{petsc-user-ref}
Balay, S., Abhyankar, S., Adams, M.~F., Benson, S., Brown, J., Brune, P.,
  et~al. (2025).
\newblock \emph{{PETSc/TAO} Users Manual}.
\newblock Tech. Rep. ANL-21/39, Argonne National Laboratory.
\newblock \doi{10.2172/2998643}
\bibAnnoteFile{petsc-user-ref}

\bibitem[{Beer et~al.(1995)Beer, Cowley, and Hammett}]{beer1995field}
Beer, M.~A., Cowley, S.~C., and Hammett, G.~W. (1995).
\newblock Field-aligned coordinates for nonlinear simulations of tokamak
  turbulence.
\newblock \emph{Phys. Plasmas} 2, 2687--2700
\bibAnnoteFile{beer1995field}

\bibitem[{Body(2022)}]{body2022development}
Body, T. A.~J. (2022).
\newblock \emph{Development of Turbulence Simulations for the Edge \& Divertor
  and Validation against Experiment}.
\newblock Ph.D. thesis, Technische Universit{\"a}t M{\"u}nchen
\bibAnnoteFile{body2022development}

\bibitem[{Cheng et~al.(2016)Cheng, Zhang, Lin, Holod, Li, Chen
  et~al.}]{cheng2016gyrokinetic}
Cheng, J., Zhang, W., Lin, Z., Holod, I., Li, D., Chen, Y., et~al. (2016).
\newblock Gyrokinetic particle simulation of fast-electron driven beta-induced
  aflv{\'e}n eigenmode.
\newblock \emph{Physics of Plasmas} 23
\bibAnnoteFile{cheng2016gyrokinetic}

\bibitem[{Daneri et~al.(2026)Daneri, Lu, Hoelzl, Greco, and
  Carr{\`a}}]{daneri2026openmp}
Daneri, G., Lu, Z.~X., Hoelzl, M., Greco, L.~V., and Carr{\`a}, E. (2026).
\newblock Openmp gpu acceleration and portability of trimeg-c1 for
  electromagnetic gyrokinetic simulations in tokamak plasmas.
\newblock \emph{arXiv preprint arXiv:2606.24327}
\bibAnnoteFile{daneri2026openmp}

\bibitem[{Fitzpatrick(2014)}]{fitzpatrick2014plasma}
Fitzpatrick, R. (2014).
\newblock \emph{Introduction to Plasma Physics} (University of Texas at
  Austin).
\newblock Lecture notes, available at
  \url{https://farside.ph.utexas.edu/teaching/plasma/Plasma/}
\bibAnnoteFile{fitzpatrick2014plasma}

\bibitem[{Hager et~al.(2022)Hager, Ku, Sharma, Chang, and
  Churchill}]{hager2022electromagnetic}
Hager, R., Ku, S.-H., Sharma, A., Chang, C., and Churchill, R. (2022).
\newblock Electromagnetic total-f algorithm for gyrokinetic particle-in-cell
  simulations with the x-point gyrokinetic code {XGC}.
\newblock \emph{arXiv preprint arXiv:2202.06124}
\bibAnnoteFile{hager2022electromagnetic}

\bibitem[{Hatzky et~al.(2019)Hatzky, Kleiber, K{\"o}nies, Mishchenko,
  Borchardt, Bottino et~al.}]{hatzky2019reduction}
Hatzky, R., Kleiber, R., K{\"o}nies, A., Mishchenko, A., Borchardt, M.,
  Bottino, A., et~al. (2019).
\newblock Reduction of the statistical error in electromagnetic gyrokinetic
  particle-in-cell simulations.
\newblock \emph{Journal of Plasma Physics} 85
\bibAnnoteFile{hatzky2019reduction}

\bibitem[{Hatzky et~al.(2002)Hatzky, Tran, K{\"o}nies, Kleiber, and
  Allfrey}]{hatzky2002energy}
Hatzky, R., Tran, T., K{\"o}nies, A., Kleiber, R., and Allfrey, S.~J. (2002).
\newblock Energy conservation in a nonlinear gyrokinetic particle-in-cell code
  for ion-temperature-gradient-driven modes in $\theta$-pinch geometry.
\newblock \emph{Physics of Plasmas} 9, 898--912
\bibAnnoteFile{hatzky2002energy}

\bibitem[{Jardin(2004)}]{jardin2004triangular}
Jardin, S.~C. (2004).
\newblock A triangular finite element with first-derivative continuity applied
  to fusion mhd applications.
\newblock \emph{Journal of Computational Physics} 200, 133--152
\bibAnnoteFile{jardin2004triangular}

\bibitem[{Kleiber et~al.(2024)Kleiber, Borchardt, Hatzky, K{\"o}nies, Leyh,
  Mishchenko et~al.}]{kleiber2024euterpe}
Kleiber, R., Borchardt, M., Hatzky, R., K{\"o}nies, A., Leyh, H., Mishchenko,
  A., et~al. (2024).
\newblock {EUTERPE}: A global gyrokinetic code for stellarator geometry.
\newblock \emph{Computer Physics Communications} 295, 109013
\bibAnnoteFile{kleiber2024euterpe}

\bibitem[{Lanti et~al.(2020)Lanti, Ohana, Tronko, Hayward-Schneider, Bottino,
  McMillan et~al.}]{lanti2020orb5}
Lanti, E., Ohana, N., Tronko, N., Hayward-Schneider, T., Bottino, A., McMillan,
  B.~F., et~al. (2020).
\newblock {ORB5}: A global electromagnetic gyrokinetic code using the {PIC}
  approach in toroidal geometry.
\newblock \emph{Computer Physics Communications} 251
\bibAnnoteFile{lanti2020orb5}

\bibitem[{Lee(1983)}]{lee1983gyrokinetic}
Lee, W.~W. (1983).
\newblock Gyrokinetic approach in particle simulation.
\newblock \emph{Phys. Fluids} 26, 556
\bibAnnoteFile{lee1983gyrokinetic}

\bibitem[{Lee(1987)}]{lee1987gyrokinetic}
Lee, W.~W. (1987).
\newblock Gyrokinetic particle simulation model.
\newblock \emph{Journal of Computational Physics} 72, 243--269
\bibAnnoteFile{lee1987gyrokinetic}

\bibitem[{Lin et~al.(1998)Lin, Hahm, Lee, Tang, and White}]{lin1998turbulent}
Lin, Z., Hahm, T.~S., Lee, W.~W., Tang, W.~M., and White, R.~B. (1998).
\newblock Turbulent transport reduction by zonal flows: Massively parallel
  simulations.
\newblock \emph{Science} 281, 1835--1837
\bibAnnoteFile{lin1998turbulent}

\bibitem[{Lu et~al.(2019)Lu, Lauber, Hayward-Schneider, Bottino, and
  Hoelzl}]{lu2019development}
Lu, Z.~X., Lauber, {\relax Ph}., Hayward-Schneider, T., Bottino, A., and
  Hoelzl, M. (2019).
\newblock Development and testing of an unstructured mesh method for whole
  plasma gyrokinetic simulations in realistic tokamak geometry.
\newblock \emph{Phys. Plasmas} 26, 122503
\bibAnnoteFile{lu2019development}

\bibitem[{Lu et~al.(2023)Lu, Meng, Hatzky, Hoelzl, and Lauber}]{lu2023full}
Lu, Z.~X., Meng, G., Hatzky, R., Hoelzl, M., and Lauber, {\relax Ph}. (2023).
\newblock Full f and $\delta$f gyrokinetic particle simulations of {Alfv\'en}
  waves and energetic particle physics.
\newblock \emph{Plasma Physics and Controlled Fusion}
\bibAnnoteFile{lu2023full}

\bibitem[{Lu et~al.(2025{\natexlab{a}})Lu, Meng, Hatzky, Lauber, and
  Hoelzl}]{lu2025trimeg}
Lu, Z.~X., Meng, G., Hatzky, R., Lauber, {\relax Ph}., and Hoelzl, M.
  (2025{\natexlab{a}}).
\newblock Trimeg-gkx: an electromagnetic gyrokinetic particle code with a
  piecewise field-aligned finite element method for micro-and macro-instability
  studies in tokamak core plasmas.
\newblock \emph{Computer Physics Communications} , 109959
\bibAnnoteFile{lu2025trimeg}

\bibitem[{Lu et~al.(2025{\natexlab{b}})Lu, Meng, Hatzky, Sonnendr{\"u}cker,
  Mishchenko, and Hoelzl}]{lu2025generalized}
Lu, Z.~X., Meng, G., Hatzky, R., Sonnendr{\"u}cker, E., Mishchenko, A., and
  Hoelzl, M. (2025{\natexlab{b}}).
\newblock Generalized mixed variable-pullback scheme with ideal/non-ideal ohm's
  law and pure symplectic scheme for electromagnetic gyrokinetic simulations.
\newblock \emph{Physics of Plasmas} 32
\bibAnnoteFile{lu2025generalized}

\bibitem[{Lu et~al.(2024)Lu, Meng, Hatzky, Sonnendr{\"u}cker, Mishchenko,
  Lauber et~al.}]{lu2024gyrokinetic}
Lu, Z.~X., Meng, G., Hatzky, R., Sonnendr{\"u}cker, E., Mishchenko, A., Lauber,
  {\relax Ph}., et~al. (2024).
\newblock Gyrokinetic electromagnetic particle simulations in triangular meshes
  with {C1} finite elements.
\newblock \emph{Plasma Physics and Controlled Fusion} 67, 015015
\bibAnnoteFile{lu2024gyrokinetic}

\bibitem[{Lu et~al.(2026)Lu, Meng, Sonnendr{\"u}cker, Hatzky, Daneri, Li
  et~al.}]{Lu2026}
Lu, Z.~X., Meng, G., Sonnendr{\"u}cker, E., Hatzky, R., Daneri, G., Li, G.,
  et~al. (2026).
\newblock A high-order piecewise field-aligned triangular finite element method
  for electromagnetic gyrokinetic particle simulations of tokamak plasmas with
  open field lines.
\newblock \emph{Plasma Physics and Controlled Fusion} 68, 055027.
\newblock \doi{10.1088/1361-6587/ae6ace}
\bibAnnoteFile{Lu2026}

\bibitem[{Lu et~al.(2025{\natexlab{c}})Lu, Meng, Sonnendr{\"u}cker, Hatzky,
  Mishchenko, Zonca et~al.}]{lu2025piecewise}
Lu, Z.~X., Meng, G., Sonnendr{\"u}cker, E., Hatzky, R., Mishchenko, A., Zonca,
  F., et~al. (2025{\natexlab{c}}).
\newblock Piecewise field-aligned finite element method for multi-mode
  nonlinear particle simulations in tokamak plasmas.
\newblock \emph{Journal of Plasma Physics} 91, E48
\bibAnnoteFile{lu2025piecewise}

\bibitem[{Michels et~al.(2021)Michels, Stegmeir, Ulbl, Jarema, and
  Jenko}]{michels2021genex}
Michels, D., Stegmeir, A., Ulbl, P., Jarema, D., and Jenko, F. (2021).
\newblock Gene-x: A full-f gyrokinetic turbulence code based on the
  flux-coordinate independent approach.
\newblock \emph{Computer Physics Communications} 264, 107986
\bibAnnoteFile{michels2021genex}

\bibitem[{Mishchenko et~al.(2019)Mishchenko, Bottino, Biancalani, Hatzky,
  Hayward-Schneider, Ohana et~al.}]{mishchenko2019pullback}
Mishchenko, A., Bottino, A., Biancalani, A., Hatzky, R., Hayward-Schneider, T.,
  Ohana, N., et~al. (2019).
\newblock Pullback scheme implementation in {ORB5}.
\newblock \emph{Computer Physics Communications} 238, 194--202
\bibAnnoteFile{mishchenko2019pullback}

\bibitem[{Mishchenko et~al.(2017)Mishchenko, Bottino, Hatzky,
  Sonnendr{\"u}cker, Kleiber, and K{\"o}nies}]{mishchenko2017mitigation}
Mishchenko, A., Bottino, A., Hatzky, R., Sonnendr{\"u}cker, E., Kleiber, R.,
  and K{\"o}nies, A. (2017).
\newblock Mitigation of the cancellation problem in the gyrokinetic
  particle-in-cell simulations of global electromagnetic modes.
\newblock \emph{Physics of Plasmas} 24
\bibAnnoteFile{mishchenko2017mitigation}

\bibitem[{Mishchenko et~al.(2014)Mishchenko, K{\"o}nies, Kleiber, and
  Cole}]{mishchenko2014pullback}
Mishchenko, A., K{\"o}nies, A., Kleiber, R., and Cole, M. (2014).
\newblock Pullback transformation in gyrokinetic electromagnetic simulations.
\newblock \emph{Physics of Plasmas} 21
\bibAnnoteFile{mishchenko2014pullback}

\bibitem[{{MPCDF}(2025)}]{viper_docs}
[Dataset] {MPCDF} (2025).
\newblock {Viper-GPU User Guide}.
\newblock
  \url{https://docs.mpcdf.mpg.de/doc/computing/viper-gpu-user-guide.html}.
\newblock Accessed: 2025-06-01
\bibAnnoteFile{viper_docs}

\bibitem[{Oliveira et~al.(2022)Oliveira, Body, Galassi, Theiler, Laribi, Tamain
  et~al.}]{oliveira2022tcvx21}
Oliveira, D.~S., Body, T., Galassi, D., Theiler, C., Laribi, E., Tamain, P.,
  et~al. (2022).
\newblock Validation of edge turbulence codes against the {TCV-X21} diverted
  {L}-mode reference case.
\newblock \emph{Nuclear Fusion} 62, 096001.
\newblock \doi{10.1088/1741-4326/ac4cde}
\bibAnnoteFile{oliveira2022tcvx21}

\bibitem[{Trott et~al.(2022)Trott, Lebrun-Grandi{\'e}, Arndt, Ciesko, Dang,
  Ellingwood et~al.}]{kokkos}
Trott, C.~R., Lebrun-Grandi{\'e}, D., Arndt, D., Ciesko, J., Dang, V.,
  Ellingwood, N., et~al. (2022).
\newblock Kokkos 3: Programming model extensions for the exascale era.
\newblock \emph{IEEE Transactions on Parallel and Distributed Systems} 33,
  805--817.
\newblock \doi{10.1109/TPDS.2021.3097283}
\bibAnnoteFile{kokkos}

\bibitem[{Ulbl et~al.(2023)Ulbl, Body, Zholobenko, Stegmeir, Pfennig, and
  Jenko}]{ulbl2023influence}
Ulbl, P., Body, T., Zholobenko, W., Stegmeir, A., Pfennig, J., and Jenko, F.
  (2023).
\newblock Influence of collisions on the validation of global gyrokinetic
  simulations in the edge and scrape-off layer of {TCV}.
\newblock \emph{Physics of Plasmas} 30
\bibAnnoteFile{ulbl2023influence}

\bibitem[{Zhang et~al.(2018)Zhang, Joubert, Wang, Wang, Tang, Niemerg
  et~al.}]{zhang2018heterogeneous}
Zhang, W., Joubert, W., Wang, P., Wang, B., Tang, W., Niemerg, M., et~al.
  (2018).
\newblock Heterogeneous programming and optimization of gyrokinetic toroidal
  code using directives.
\newblock In \emph{International Workshop on Accelerator Programming Using
  Directives} (Springer), 3--21
\bibAnnoteFile{zhang2018heterogeneous}

\end{thebibliography}

\end{document}